\documentclass[prc, showpacs, letterpaper, preprintnumbers,amsmath,amssymb]{revtex4}
\usepackage{amsmath}
\usepackage{amssymb}
\usepackage{color}
\usepackage{mathrsfs}
\usepackage{graphics}
\usepackage{graphicx}
\usepackage{dcolumn}
\usepackage{bm}

\newcommand{\lsim}{\mathrel{\raisebox{-.6ex}{$\stackrel{\textstyle<}{\sim}$}}}
\newcommand{\gsim}{\mathrel{\raisebox{-.6ex}{$\stackrel{\textstyle>}{\sim}$}}}
\newcommand{\slsh}{\mbox{ \hspace{-1.1 em} $/$}}
\newcommand{\slst}{\mbox{ \hspace{-1.5 em} $/$}}

\def\ni{\noindent}
\begin{document}
\title{The photo-neutrino process in astrophysical systems}

\author{Sharada Iyer Dutta}
\email{iyers@neutrino.ess.sunysb.edu}

\author{Sa{\v s}a Ratkovi\'c}
\email{ratkovic@tonic.physics.sunysb.edu}

\author{Madappa Prakash}
\email{prakash@snare.physics.sunysb.edu}

\affiliation{Department of Physics \& Astronomy, \\
        State University of New York at Stony Brook, \\
        Stony Brook, NY 11794-3800, USA}


\date{\today}

\begin{abstract}

Explicit expressions for the differential and total rates and
emissivities of neutrino pairs from the photo-neutrino process $e^\pm
+ \gamma \rightarrow e^\pm + \nu + \bar\nu$ in hot and dense matter
are derived.  Full information about the emitted neutrinos is retained
by evaluating the squared matrix elements for this process which was
hitherto bypassed through the use of Lenard's identity in obtaining
the total neutrino emissivities.  Accurate numerical results are
presented for widely varying conditions of temperature and density.
Analytical results helpful in understanding the qualitative behaviors
of the rates and emissivities in limiting situations are derived.  The
corresponding production and absorption kernels in the source term of
the Boltzmann equation for neutrino transport are developed.  The
appropriate Legendre coefficients of these kernels, in forms suitable
for multigroup flux-limited diffusion schemes are also provided.

\end{abstract}
\smallskip
\maketitle


\section{Introduction}
\label{sec:Introduction}

In recent years, the study of neutrino emission, scattering, and
absorption in matter at high density and/or temperature has gained
prominence largely due to its importance in a wide range of
astrophysical phenomena.  Energy loss in degenerate helium cores of
red giant stars \cite{Gross,Raffelt(2000)}, cooling in pre-white dwarf
interiors \cite{O'BrienKawaler(2000)}, the short- and long-term
cooling of neutron stars \cite{Prak02,Yak02}, the deflagration stages
of white dwarfs which may lead to type Ia supernovae
\cite{Wolfgang,Iwamoto}, explosive stages of type II (core-collapse)
supernovae \cite{Bur00}, and thermal emission in accretion disks of
gamma-ray bursters \cite{Matteo,Kohri}, are examples in which neutral
and charged current weak interaction processes that involve neutrinos
play a significant role.  (The selected references contain more
complete references to prior and ongoing work.)

In unravelling the mechanism by which a type-II supernova explodes,
the implementation of accurate neutrino transport has been realized to
be critical \cite{Trans}.  The basic microphysical inputs of accurate
neutrino transport coupled in hydrodynamical situations are the
differential neutrino production and absorption rates and their
associated emissivities.  The processes and precise forms in which
such inputs are required for multienergy treatment of neutrinos for
both sub-nuclear and super-nuclear densities (nuclear density $\rho_0
\simeq 2.65 \times 10^{14}~{\rm g~cm^{-3}})$ are detailed in
Refs. \cite{BRUENN1,BT02}. At sub-nuclear densities, detailed
differential information is available for pair production ($e^+ + e^-
\rightarrow \nu + \bar\nu$) \cite{BRUENN1}, nucleon bremsstrahlung ($n
+ n \rightarrow n + n + \nu + \bar\nu$) \cite{HR98}, $\nu$-flavor
production ($\nu_i + \bar{\nu}_i \rightarrow \nu_j + \bar{\nu}_j$)
\cite{Buras02}, and more recently for the plasma process ($\gamma^*
\rightarrow \nu +\bar\nu$) \cite{RIP02}.

Our objectives in this work are to make available differential rates,
emissivities, and the source and sink terms associated with the
thermal photo production of neutrino-pairs in a Boltzmann transport
treatment of the process $e^\pm + \gamma \rightarrow e^\pm + \nu + \bar\nu$
for which only the total rates and emissivities are available to
date \cite{BEAUDET1,BEAUDET2,DICUS1,BOND1,SCHINDER1,ITOH}.  (It is
important to note that in prior works, the energy and angular
dependences of the emitted neutrinos were lost in simplifying the
calculations of the total rates and emissivities; see Sec. II for
details).  In addition, we provide a qualitative physical
understanding of the behavior of the neutrino emissivity for 
widely varying conditions of density and temperature.

Section \ref {sec:Photo} is devoted to the derivation of working
expressions for the differential rates and emissivities from the
photo-neutrino process.  The calculation of the hitherto unavailable
squared matrix elements is outlined in Sec. II. A. Explicit
expressions of these matrix elements, including those for the
transverse and longitudinal components, are given in Appendix A.  In
Sec. II. B, the expressions for rates and emissivities are rendered in
a form suitable for numerical calculations.  The input photon
dispersion relation is briefly discussed in Sec. II. C. Notes for
obtaining accurate results from Monte Carlo integrations of the
8-dimensional integrals are provided in Sec. II. D and Appendix B
where the resolution of numerical problems encountered in earlier
works is addressed. Results of
numerical calculations are presented in Sec.  III. The subsections
here contain an analytical analysis of the qualitative behaviors of
the total rates and emissivities for widely varying conditions of
density and temperature.  Sec. \ref{sec:PKER} details the derivation
of the production and absorption kernels in forms suitable for
detailed calculations of neutrino transport along with numerical
results for the leading Legendre coefficients. Sec. V contains a
summary and a discussion of the relation of this work with those of
prior works.  Except when presenting numerical results, we use units
in which $\hbar,~c,~{\rm and }~k_B$ are set to unity.
\begin{figure}[h!]
\begin{picture}(250, 200)(0, 0)
\put(000,000){\includegraphics[width=100pt]{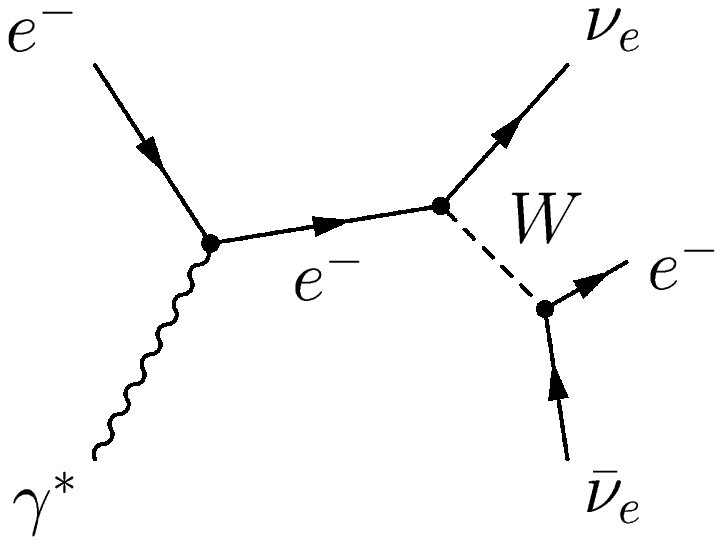}}
\put(125,000){\includegraphics[width=100pt]{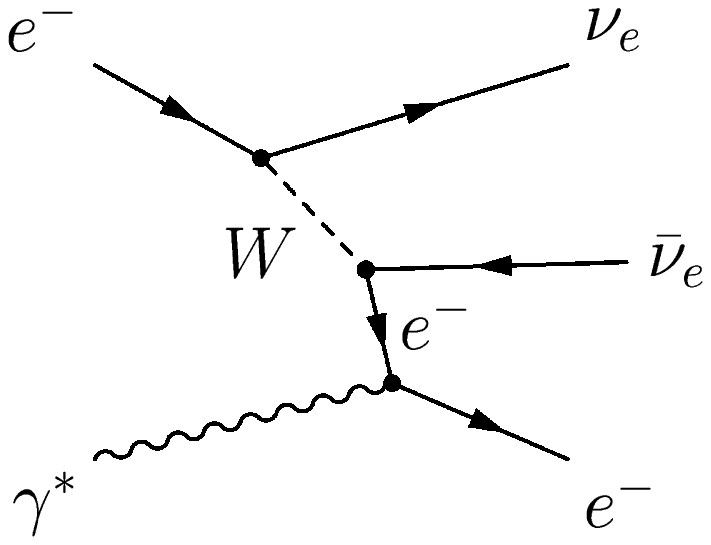}}
\put(000,100){\includegraphics[width=100pt]{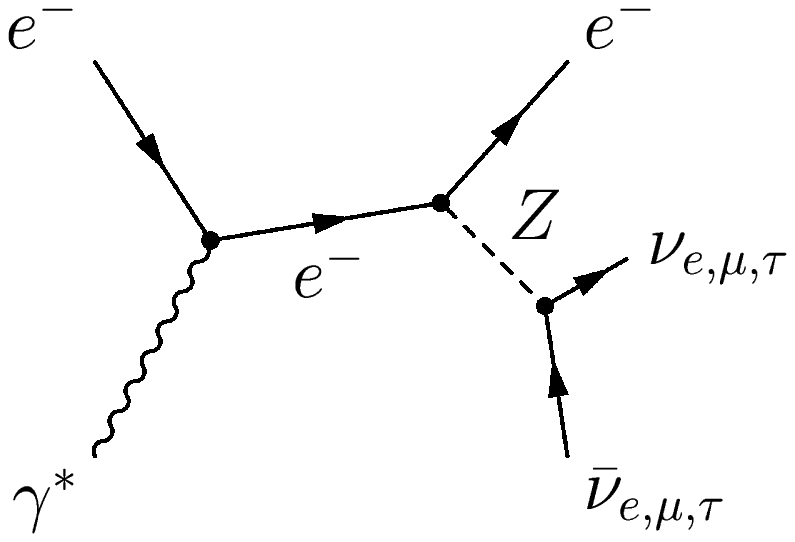}}
\put(125,100){\includegraphics[width=100pt]{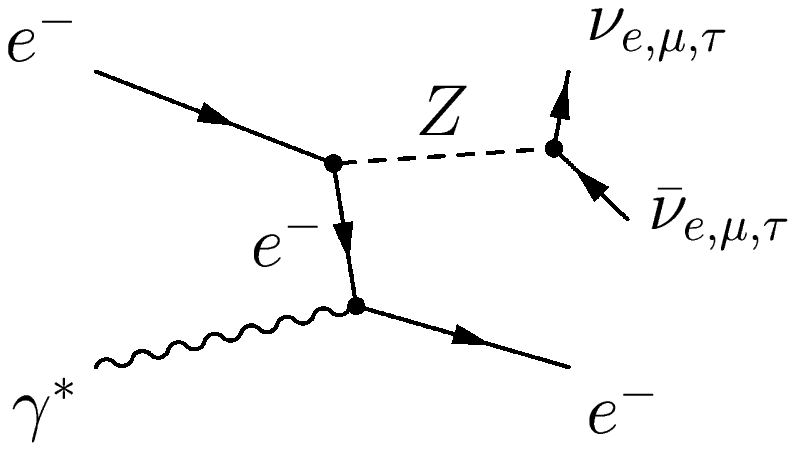}}
\end{picture}%
\caption{Leading order Feynman diagrams describing the emission of a
neutrino pair from the photo process.  The charged current $W$ -
exchange channel produces only $\nu_e \bar{\nu}_e$, whereas the
neutral $Z$ - exchange results in pairs of all three neutrino
($e,~\mu,~{\rm and}~\tau$) flavors. Contributions from positrons are
obtained by the replacement $e^- \rightarrow e^+$.}
\label{photo}
\end{figure}


\section{The Photo-neutrino Process}
\label{sec:Photo}

The leading order diagrams for the photoproduction of neutrino pairs,
$e^\pm + \gamma\rightarrow e^\pm + \nu_{e\,(\mu,\tau)} +
\bar{\nu}_{e\,(\mu,\tau)}$  are shown in Fig. \ref {photo}. The
channel in which the exchange of the $Z$-boson occurs can produce any
of the three species of neutrinos ($\nu_{e,\mu,\tau}$) and their
anti-particles, whereas the channel in which the $W$-boson is
exchanged, only the $\nu_e \bar \nu_e$ pair is produced.

The total emissivity or the total energy carried away by the neutrino
pair per unit volume per unit time from the photo-neutrino process is
\begin{eqnarray}
\nonumber
Q&=&\int\frac{2\,d^3{\bf p}}{(2\pi)^3} \frac {F_e(E_p)}{2E_p} 
\int\frac{\xi \,d^3{\bf k}}{(2\pi)^3} \frac {F_{\gamma}(\omega)}{2\omega} 
\int\frac{d^3{\bf p^{\prime}}}{(2\pi)^3} 
\frac{ [1-F_e(E_{p^{\prime}})]}{2E_{p^{\prime}}}
\int\frac{d^3{\bf q}}{(2\pi)^3} \frac{1}{2E_q}
\int\frac{d^3{\bf q^{\prime}}}{(2\pi)^3}  \frac{1}{2E_{q^{\prime}}} \\ &&\times
\,\, (2\pi)^4 \delta^4(p+k-p^{\prime}-q-q^{\prime})(E_q+E_{q^{\prime}})
\frac{1}{\zeta} \sum_{s,\epsilon}|{\mathcal{M}}|^2 .
\label{Emisstot}
\end{eqnarray}
The first factor 2 accounts for the spin projections of the incoming
electron.  The factor $\xi$ accounts for the polarizations of the
incoming photon and the factor $\zeta$ arises from averaging over the
spin projections of the outgoing electron and neutrinos.  For the
transverse polarization of the photon, $\xi = 2$ and $\zeta = 4$,
while for the longitudinal polarization $\xi = 1$ and $\zeta = 2$. The
index $s$ keeps track of the spin projections of the initial and final
electrons, while the index $\epsilon$ does the same for the
polarization states of the photon.  The four-momenta of the
participating particles are:
\begin{eqnarray}
p  &\equiv& (E_p,{\bf p}): \hspace*{1.5cm} {\rm ~incoming~electron\,,}
\nonumber \\
p^{\prime} &\equiv& (E_{p^{\prime}},{\bf p}^\prime): 
\hspace*{1.4cm} {\rm outgoing~electron\,,}
\nonumber \\
k  &\equiv& (\omega,{\bf k}): 
\hspace*{1.6cm}{\rm ~~incoming~in-medium~massive~photon\,,} 
\nonumber \\
 q &\equiv& (E_q,{\bf q}): \hspace*{1.6cm} {\rm outgoing~neutrino\,,~~and}
\nonumber \\ 
 q^{\prime}&\equiv& (E_{q^{\prime}},{\bf q}^\prime): 
\hspace*{1.4cm} {\rm outgoing~anti-neutrino\,,}
\label{4mom} 
\end{eqnarray}
where the energies and three-momenta are indicated in standard
notation. In the entrance channel, electrons and photons are drawn
from equilibrium Fermi-Dirac and Boson-Einstein distribution functions 
\begin{eqnarray}  
F_e(E_p) &=& \left[\exp \left(\frac{E_e - \mu_e}{T} \right) +
1\right]^{-1}\,, \hspace{1cm}
E_p = {\sqrt {{\bf|p|}^2 + m_e^2}} \,, \quad {\rm and} \nonumber \\ 
F_\gamma(\omega ) &=& \left[\exp \left(\frac{\omega}{T} \right) -
1\right]^{-1} \,,  \hspace{2.2cm}  \omega =  \omega ({\bf k}) = 
{\sqrt {{\bf |k|}^2 + \omega_p^2}} + \cdots \,,    
\end{eqnarray}
respectively. The quantities $\mu_e$ and $T$ denote the electron
chemical potential and temperature, while $m_e$ and $\omega_p$ stand
for the electron mass and plasma frequency (this is related to the
in-medium photon mass), respectively. The factor
$1-F_e(E_{p^{\prime}})$ accounts for the Pauli-blocking of the
outgoing electrons.  Blocking factors are absent for neutrinos, since
the emitted (low-energy) neutrinos leave the production site without
further interactions.

The rate or number of neutrino pairs produced per unit volume per unit time,
$\Gamma$, is given by an expression analogous to that in
Eq. (\ref{Emisstot}), but without the factor $E_q+E_{q^\prime}$ in the
integrand.  In prior works in which the total rates and emissivities
were computed, the energy and angular dependences of the emitted
neutrinos were eliminated by using Lenard's identity~\cite{Lenard}:
\begin{equation}
\int \frac {d^3q}{2E_q}   \frac {d^3q^{\prime}}{2E_{q^{\prime}}} 
\delta^4(q_t-q-q^{\prime} ) q^\mu q{^{\prime\nu}} = \frac {\pi}{24} 
\Theta(q_t^2) (2 q_t^\mu q_t^\nu + q_t^2 g^{\mu\nu})  \,,  
\label{Lenard}
\end{equation}  
where $q_t = q + q^{\prime} = p + k - p^{\prime}$.  Although the use
of this identity simplifies considerably the calculation of the total
emissivity, differential information about the neutrinos is entirely
lost.  On the other hand, calculations of differential rates and
emissivities, such as
\begin{equation}
\frac {d^3\Gamma}{dE_q\, dE_{q^\prime}\, d(\cos\theta_{qq^{\prime}})} \, 
\qquad {\rm and} \qquad 
\frac {d^3Q}{dE_q\, dE_{q^\prime}\, d(\cos\theta_{qq^{\prime}})} \, , 
\label{diffs}
\end{equation} 
where $\theta_{qq^{\prime}}$ is the angle between the neutrino pairs,
entail the calculation of the relevant squared matrix element 
hitherto bypassed in obtaining the total rates and emissivities. We
therefore turn now to the evaluation of the squared matrix element.

\subsection{Matrix Elements}
\label{subsec:EPC}

The $Z-$ and $W-$ exchange contributions to the total matrix element
${\mathcal {M}} = {\mathcal {M}}_Z+{\mathcal {M}}_W$ can be combined
by a Fierz transformation to yield \cite{DICUS1}
\begin{equation}
{\mathcal{M}} = -\frac{ieG_F}{\sqrt 2}
\bar{u}_e(p^{\prime}) \left[\gamma^\alpha(C_V-C_A\gamma_5)
\frac{p \slsh +k \slsh +m_e}{2p\cdot k+k^2}\epsilon \slsh  
+ \epsilon \slsh 
\frac{p^{\prime} \slst -k \slsh +m_e}{-2p^{\prime}\cdot k+k^2}
\gamma^\alpha(C_V-C_A\gamma_5)\right]u_e(p)~
\bar{u}_\nu(q)\,\gamma_\alpha\,(1-\gamma_5)\,u_{\bar{\nu}}\,(q^{\prime}) \,,
\label{Matrix}
\end{equation}
where $e$ is the charge of the electron and $G_F$ is the weak (Fermi)
coupling constant.  For the $e^\pm + \gamma\rightarrow e^\pm + \nu_e
+ \bar{\nu}_e$ process, the numerical values of the vector and axial couplings,
$C_V$ and $C_A$, are
\begin{eqnarray}
\nonumber
C_V = \frac{1}{2}+2\,\sin^2\theta_W  \qquad  {\rm and} \qquad 
C_A  = \frac{1}{2}  \,,
\label{Cconstant}
\end{eqnarray}
where $\sin^2\theta_W=0.226$. 
The spin-summed squared matrix element takes the form
\begin{eqnarray}
\nonumber
\sum_s |{\mathcal{M}}|^2 \, &=& \frac{e^2G_F^2}{2}
{\rm Tr} \biggl((p^{\prime} \slst +m_e)\biggl[\gamma^{\alpha}
(C_V-C_A\gamma_5)\frac{(Q
\slsh_1+m_e)}{\beta_1}\epsilon \slsh +\epsilon \slsh \frac{(Q
\slsh_2+m_e)}{\beta_2}\gamma^{\alpha}(C_V-C_A\gamma_5)\biggr](p
\slsh+m_e)\\ &&\biggl[(C_V+C_A\gamma_5)\gamma^{\beta}\frac{(Q
\slsh_2+m_e)}{\beta_2}\epsilon \slsh +\epsilon \slsh \frac{(Q
\slsh_1+m_e)}{\beta_1}(C_V+C_A\gamma_5)\gamma^{\beta}\biggr]\biggr)\,\,
{\rm Tr} \biggl(q\slsh\gamma_\alpha(1-\gamma_5)(q\slsh^{\prime})(1+\gamma_5)
\gamma_\beta\biggr) \,,
\label{Matsqr}
\end{eqnarray}
where we have introduced the symbols 
\begin{eqnarray}
Q_1 &=& p+k\,, \hspace{2.5cm}   Q_2=p^{\prime}-k \,, \nonumber \\
\beta_1 &=& 2p\cdot k + k^2 \qquad {\rm and} \qquad 
\beta_2  = - 2 p^{\prime}\cdot k + k^2 \,.
\end{eqnarray} 
The emissivity from each of the $e^\pm +
\gamma\rightarrow e^\pm + \nu_{(\mu,\tau)} + \bar{\nu}_{(\mu,\tau)}$
processes is obtained by the replacements
\begin{eqnarray}
C_V \rightarrow C_V -1 \qquad {\rm and} 
\qquad   C_A \rightarrow C_A - 1 \,. 
\end{eqnarray}
The total emissivity of all three neutrino flavors is obtained by the
replacements
\begin{eqnarray}
C_V ^2 \rightarrow C_V ^2 +2(C_V -1)^2 \qquad {\rm and} \qquad 
C_V ^2 \rightarrow C_A ^2 +2(C_A -1)^2\,. 
\end{eqnarray}
For massless neutrinos, the trace over the outgoing neutrinos yields
the familiar tensor
\begin{eqnarray}
L_{\alpha\beta}&=&8\biggl[(q_\alpha q^{\prime}_\beta +q^{\prime}_\alpha q_\beta
-g_{\alpha\beta}\,\,q\cdot q^{\prime}) + 
i\epsilon_{\mu\alpha\nu\beta}\,\,q^{\mu}q^{\prime\nu}\biggr].
\label{lepten}
\end{eqnarray}
It remains then to contract this neutrino tensor with that obtained by
performing the trace over terms that couple the electron with the
photon.  The result can be expressed as 
\begin{eqnarray}
\sum_{s,\epsilon} | {\mathcal{M}}|^2 &=& {32e^2G_F^2}~
\,\,\sum_\epsilon \biggl\{(C_V^2-C_A^2)\,\,m_e^2\,\,\mathcal{M_-}+
(C_V^2+C_A^2) \,\,\mathcal{M_+} + 
C_V\,\,C_A \,\,\mathcal{M_\times}\biggr\} \,,
\label{ME2}
\end{eqnarray}
where the quantities $ \mathcal{M_-},\,\mathcal{M_+}$ and
$\mathcal{M_\times}$ depend on the scalar products of the various four
momenta and the polarization of the photon.  The remaining sum over
the photon polarizations is performed in terms of its longitudinal
and transverse components:
\begin{eqnarray}
\sum_{\lambda=1}^3{\epsilon^{*(\lambda)}}^{\mu}{\epsilon^{(\lambda)}}^\nu 
= - g^{\mu\nu} + \frac {k^\mu k^\nu}{k^2} = P_T^{\mu\nu}+P_L^{\mu\nu}\, .
\end{eqnarray}
The components of the transverse and longitudinal polarization tensors are 
\begin{eqnarray}
P_T^{\mu\nu}&=&\left\{ \begin{array}{ll}
      0& \textrm{for } \mu \textrm{ or }\nu=0\\
      \delta^{ij}-\frac{k^ik^j}{k^2}&i,j=1, 2, 3
\end{array}
\right. \, ,
\end{eqnarray}
\begin{eqnarray}
P_L^{\mu\nu}&=&-g^{\mu\nu}+\frac{k^\mu k^\nu}{k^2}-P_T^{\mu\nu} \,.
\end{eqnarray}
These polarization tensors satisfy the properties
\begin{eqnarray}
P_T^{\mu\rho}P_{L\rho\nu}&=&0 \,, \quad
P_T^{\mu\rho}P_{T\rho\nu} = -P_{T\nu}^\mu \nonumber\\
P_L^{\mu\rho}P_{L\rho\nu}&=&-P_{L\nu}^\mu\,, \quad 
P_{L\mu}^\mu = -1 \,, \quad 
P_{T\mu}^\mu = -2 \,.
\end{eqnarray}
Explicitly, the transverse and the longitudinal components of the
squared matrix element are
\begin{eqnarray}
\sum_{s,\epsilon} |\mathcal{M}^{T\,(L)}|^2  &=&  {32e^2G_F^2}~
\,\,\biggl\{(C_V^2-C_A^2)\,\,m_e^2\,\,{\mathcal{M_-}}^{T\,(L)}+
(C_V^2+C_A^2)\,\,{\mathcal{M_+}}^{T\,(L)} + 
C_V\,\,C_A\,\,{\mathcal{M_\times}}^{T\,(L)}
\biggr\} \,,
\label{ME2TL}
\end{eqnarray}
Expressions for the quantities $ \mathcal{M_-},\,\mathcal{M_+}$ 
and $\mathcal{M_\times}$, and their transverse and longitudinal components 
are given in Appendix \ref{sec:smes}.

\subsection{Differential and Total Emissivities}
\label{sec:DIFFEMIS}

In this section, we derive expressions for the differential and total
emissivities in forms that are suitable for numerical
calculations. Similar quantities for the rates can be calculated
analogously by dropping the factor $E_q+E_{q^{\prime}}$ in the
integrand. We begin by rewriting the expression for the total
emissivity in Eq.~(\ref{Emisstot}) as

\begin{eqnarray}
 Q &=& \frac{1}{(2\pi)^9}
\int\frac{d^3{\bf q}}{ 2E_q}
\int\frac{d^3{\bf q^\prime}}{ 2E_{q^\prime}}
\int\frac{d^3{\bf p^\prime}}{ 2E_{p^\prime}}
~[1-F_e(E_{p^\prime})]~(E_q+E_{q^\prime})
\,\,I({p^\prime},q, {q^\prime}) \\
I(p^\prime,q,q^\prime) &=& \frac{1}{(2\pi)^2}
\int\frac{d^3{\bf p}}{2E_p}\,\,\int
\frac{d^3{\bf k}}{2\omega}~F_{\gamma}(\omega)\,\,F_e(E_p)\,\,
\delta^4(p+k-P) \,\, \sum_{s,\epsilon} |\mathcal{M}^{T\,(L)}|^2 \,, 
\label{Int1}
\end{eqnarray}
where the total four momentum and invariant squared mass are denoted
by
\begin{equation}
P=(E,\,{\bf P})= p+k=p^\prime+q+q^\prime \,\,\quad \,\,
{\rm and} \qquad P^2=M^2\, .
\label{InvP}
\end{equation}
Note that the quantity $I(p^\prime,q, q^\prime)$ involves integrations over the
incoming particles only.  Utilizing the three-momentum delta function
to integrate over the momentum ${\bf p}$ of the incoming electron, we
obtain
\begin{eqnarray}
I(p^\prime,q,q^\prime)=\frac{1}{4(2\pi)^2}
\int_0^\infty \frac{{\bf |k|}^2}{E_p\,\omega} d{\bf |k|}\int_0^{2\pi} 
d\phi_k\int_{-1}^1 d(\cos\,\theta_k)\,
F_{\gamma}(\omega)\,\,F_e(E_p)\,\,
\delta(E-E_p-\omega) \,\, \sum_{s,\epsilon} |\mathcal{M}^{T\,(L)}|^2\, . 
\label{Int2}
\end{eqnarray}
The energy delta function can be employed to perform integration over
the angle ${\theta_k}$ between ${\bf P}$ and $ {\bf k}$ by using
\begin{eqnarray}
\frac{\delta(E-E_p-\omega)}{2E_p}
&=&\frac{1}{2\,{\bf |P|\,|k|}}
\delta\biggl(\cos\,\theta_k - \frac{m_e^2+{\bf|P|}^2+{\bf |k|}^2-(E-\omega)^2}
{2\,{\bf |P|\,|k|}}\biggr)
\label{Int3}
\end{eqnarray}
which sets
\begin{eqnarray}
\cos\,\theta_k &=&\frac{m_e^2-M^2-m_k^2+2E\omega}
{2\,{\bf |P|\,|k|}}\, ,
\label{Int4}
\end{eqnarray}
where $k^2=m_k^2 = \omega^2- {\bf |k|}^2$.  Integration over $\theta_k$ 
yields
\begin{eqnarray}
I(p^\prime,q,q^\prime)=\frac{1}{4(2\pi)^2}
\int_0^\infty  \frac{{\bf |k|}}{\omega} d{\bf|k|}\int_0^{2\pi} 
d\phi_k\,\,
F_{\gamma}(\omega)\,\,F_e(E_p)\,\frac{1}{{\bf |P|}} 
\,\, \sum_{s,\epsilon} |\mathcal{M}^{T\,(L)}|^2 \,.
\label{Int5}
\end{eqnarray}
The condition $|\cos \theta_k|\le 1$ combined with Eq. (\ref{Int4}) can
be used to establish the range in which either the energy or momentum of
the incoming photon is able to conserve the total four-momentum
$P$. However, the appropriate choice of the integration variable is
dictated by the precise form of the dispersion relation $\omega =
{\sqrt {{\bf |k|}^2 + m_k^2}}$.  In the case that the mass of the
photon is independent of its momentum, the momentum integration in
Eq. (\ref{Int5}) can be swapped with an energy integration using
${\bf|k|}\,d{\bf |k|} = \omega\,d\omega$.  The approximate dispersion
relation, $\omega = {\sqrt {{\bf |k|}^2 + \omega_p^2}}$, for the
transverse photon is an example of this situation in which
$m_k=\omega_p$.  In this case, the photon energy must satisfy
\begin{eqnarray} 
M^2\,\omega^2+E\,\omega\,(m_e^2-m_k^2-M^2)+\frac{(m_e-m_k^2-M^2)^2}{4}+m_k^2
{\bf |P|}^2\,\,\le \,\,0\, .
\label{Int6}
\end{eqnarray}
The roots of the above quadratic equation 
\begin{eqnarray}
\omega^{\pm} &=& \frac{E(M^2+m_k^2-m_e^2)\pm {\bf |P|}\sqrt{(M^2+m_k^2-m_e^2)
^2-4M^2m_k^2}}{2M^2}
\label{Int7}
\end{eqnarray}
specify the range in which the photon energy must lie.

In the more general case that $m_k$ depends on the photon momentum
${\bf |k|}$, it is convenient to retain momentum as the integration
variable in Eq.~(\ref{Int5}).  In this case, the range of momentum
integration is obtained by the solutions of
\begin{eqnarray}
\label{eq:p1limits}
	M^2 {{\bf |k|}^\pm}^2 - 
	E(M^2+m_k^2-m_e^2) \sqrt{{{\bf |k|}^\pm}^2+m_k^2}
	+ {E}^2 m_k^2 + \frac{1}{4}(M^2+m_k^2-m_e^2)^2
	=0
\end{eqnarray}
which can be found by using an iterative procedure. 

In computing the differential and total emissivities or rates, it is
desirable to allow the energies and the angle between the outgoing
neutrinos to be as unrestricted as possible.  The choice of coordinate
axes that accomplishes this requirement for the outgoing particles and
the concomitant restrictions on the incoming particles are described below.

\subsubsection*{Outgoing Particles} 
\begin{itemize}
\item The 3-momentum of one of the neutrinos is aligned along the 
z-axis.  This fixes its 4-momentum to be
\begin{eqnarray}
q=(E_q,\,{\bf q}
)&=&(E_q, \,0,\,\,0,\,\,|{\bf q}|)=(E_q,\, 0,\,\,0,\,\,E_q)\, .
\end{eqnarray}
\item The other neutrino  is restricted to the x-z plane so
that its 4-momentum is 
\begin{eqnarray}
q^\prime=(E_{q^\prime},\,{\bf q}^\prime)&=&
(E_{q^\prime},\,|{\bf q^\prime}|\, \sin\,\theta_{qq\prime}\,,\,0,\,
|{\bf q^\prime}|\,\cos\,
\theta_{qq\prime})=
(E_{q^\prime},\,E_{q^\prime}\, \sin\,\,\theta_{qq\prime}\,,\,0,\,
E_{q^\prime}\,\cos\,\theta_{qq\prime})\, ,
\end{eqnarray}
where $\theta_{qq\prime} \,\, \epsilon \,\,[0, \pi]$ is the angle
between the two neutrinos.
\item The energy  $E_{p^\prime}$ and the angles  $\theta_e
\,\,\epsilon\,\, [0, \pi]$, and $\phi_e \,\,\epsilon \,\,[0, 2\pi]$ of the
outgoing electron vary unrestricted in their domain so that 
\begin{eqnarray}
p^\prime=(E_{p^\prime},\,{\bf p}^\prime)&=&(E_{p^\prime},\,{\bf |p^\prime|}
 \,\sin\,\theta_e\,\cos\,\phi_{e}\,,\,{\bf |p^\prime|} \,
\sin\,\theta_e \,\sin\,\phi_{e}\,,\,{\bf
 |p^\prime|}\,\cos\,\theta_e)\, .
\end{eqnarray}
\end{itemize}

\ni The specification of the outgoing momenta enables the
determination of the total 4-momentum $P$, through Eq.~(\ref{InvP}) and hence $|{\bf P}|$ and
angles $\theta_P$, $\phi_P$ can be obtained.

\subsubsection*{Incoming Particles}
\begin{itemize}
\item Once $P$ is known, the integration limits $|{\bf k}|^\pm$ for
the photon momentum can be determined by using
Eq.~(\ref{eq:p1limits}).  The simplest way to enforce these
restrictions on $|{\bf k}|^\pm$ is to use a frame in which
${\bf P}$ is oriented along the $\hat z$ axis.  For $|{\bf k}|$ to be
in the allowed range in this frame, $\theta_k$ is set from
Eq.~(\ref{Int4}) and $\phi_k$ lies between $0$ and $2\pi$.  If energy
is chosen to be the integration variable, Eq.~(\ref{Int7}) provides
the appropriate range for the photon energy.

\item In order to use the newly determined  incoming photon and electron
momenta, we have to transform them back to the coordinate system
determined by the outgoing neutrino-antineutrino pair.  This is easily
achieved by a rotation of ${\bf k}$ into the frame of the outgoing
particles. The rotation is performed in two successive steps: first a
positive rotation by $\theta_P$ about the x axis and then a subsequent
rotation by $\phi_p-3\pi/2$ about the z-axis.

\item Finally, the 4-momentum of the incident electron $p$ is found
from $p=P-k$.
\end{itemize}
The coordinate axes being chosen, the 
working expression for the total emissivity takes the form   
\begin{eqnarray}
Q=\frac{\pi^2}{(2\pi)^9}
\int_0^\infty E_q\,dE_q\int_0^\infty E_{q^\prime}\,dE_{q^\prime}
\int_{-1}^{1} d\cos\theta_{qq\prime}
\int_0^\infty \frac{{\bf |p^\prime|}^{2}}
{E_{p^\prime}}d{\bf |p|}^\prime \int_{-1}^{1}
 d\cos\theta_{e}\int_0^{2\pi} d\phi_{e}[1-F_e(E_{p^\prime})]
(E_q+E_{q^\prime})I(p^\prime,q,q^\prime)\, , \nonumber\\
\label{Int8}
\end{eqnarray}
where the quantity $I(p^\prime,q, q^\prime)$ is given in Eq.~(\ref{Int5}). 
It is now straightforward to read off the differential emissivity from
this result. Explicitly, 
\begin{eqnarray}
\frac {d^3Q}{dE_q\, dE_{q^\prime}\, d(\cos\theta_{qq^{\prime}})}  = 
\frac{\pi^2}{(2\pi)^9}
E_q E_{q^\prime}\,
\int_0^\infty \frac{{\bf |p^\prime|}^{2}}
{E_{p^\prime}}d{\bf |p|}^\prime \int_{-1}^{1}
 d\cos\,\theta_{e}\int_0^{2\pi} d\phi_{e}[1-F_e(E_{p^\prime})]
(E_q+E_{q^\prime})
\,\,I(p^\prime,q, q^\prime)\, .
\label{dInt1}
\end{eqnarray}
\subsection{The Photon Dispersion Relation}
\begin{figure}
\includegraphics[width = 0.5\textwidth]{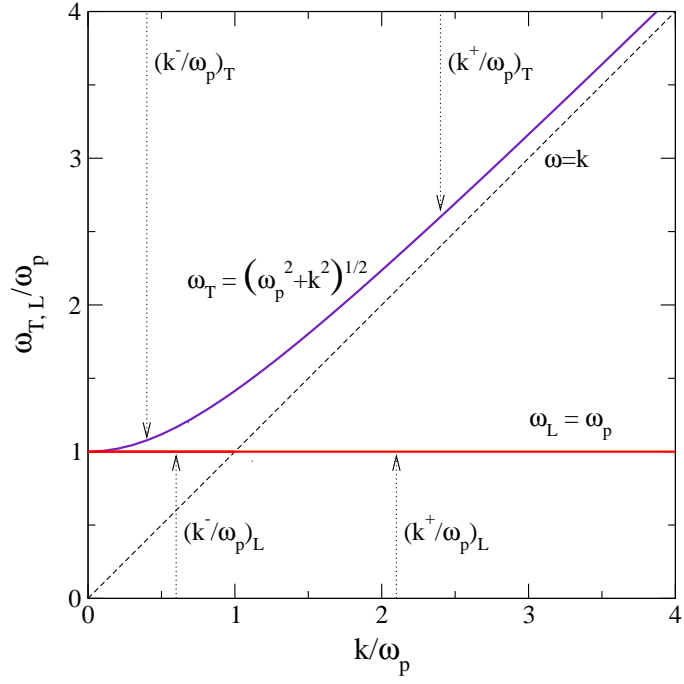}
\caption{The leading order transverse and longitudinal 
photon dispersion relations $\omega_T$ and $\omega_L$ used in the numerical
evaluation of the neutrino emissivities. The momentum cut-offs
$(k^{\pm}/\omega_p)_{T,L}$ depend on the density and temperature of
the plasma.} 
\label{fig:approxdisp}
\end{figure}

The dispersion relations of the photon in a plasma are commonly
written as
\begin{eqnarray}
\omega_T^2(k) &=& k^2 + \Pi_T(\omega_T(k),k): \hspace*{1.0cm} {\rm Transverse}
\nonumber \\ 
\omega_L^2(k) &=& \frac {\omega_L^2(k)}{k^2} \Pi_L(\omega_L(k),k): 
\hspace*{0.8cm} {\rm Longitudinal}\, , 
\end{eqnarray}
where the transverse and longitudinal polarization functions $\Pi_T$
and $\Pi_L$ account for the effects of the medium on the photon.  An
extensive analysis of the exact dispersion relations for the densities
and temperatures of interest here can be found in Refs. \cite{RIP02,
BRAATEN1}. Generally, calculations of $\Pi_T$ and $\Pi_L$ involve
iterative procedures to solve either transcendental algebraic or
integral equations.  The transverse or longitudinal emissivities are
then obtained by setting $m_k^T =\sqrt{\Pi_T(k)}$ or $m_k^L =
\sqrt{\omega_L^2-{\bf |k|}^2}$, respectively, with the integration limits on
the photon momentum ${\bf |k|}^\pm$ obtained iteratively from
Eq. (\ref{eq:p1limits}).  This procedure, although straightforward, is
time consuming. 
Fortunately, in the temperature and density ranges in which the
photo-neutrino process dominates over the other competing processes, 
the leading order dispersion relations (see Fig. \ref{fig:approxdisp}) 
\begin{eqnarray}
\label{eq:tdispersion}
\omega_T^2 &=& \omega_p^2+{\bf|k|}^2\,: \hspace*{1.1cm} {\rm Transverse} \\
\omega_L^2 &=& \omega_p^2\,:  \hspace*{2.1cm} {\rm Longitudinal}\, 
\label{eq:ldispersion}
\end{eqnarray}
give an accurate representation of the full
results.  For the most part, therefore, we will present results by
using these dispersion relations both because it is computationally
faster and because it affords comparisons with results of earlier
work.

The relations in Eqs. (\ref{eq:tdispersion}) and
(\ref{eq:ldispersion}) yield
\begin{eqnarray}
\label{eq:tphotonmass}
m_k^T &=& \omega_p\,: \hspace*{2.3cm} {\rm Transverse} \\
m_k^L &=& \sqrt{\omega_p^2-{\bf|k|}^2}\,: \hspace*{1cm} {\rm
Longitudinal}\, . 
\label{eq:lphotonmass}
\end{eqnarray}
For the transverse photon, the momentum cutoffs are given by  
\begin{eqnarray}
|{\bf k}|^\pm&=&\biggl[\biggl(\frac{E A\pm|{\bf \bar P}|
\sqrt{A^2-4M^2\omega_p^2}}{2M^2}\biggl)^2-\omega_p^2
\biggr]^{1/2}\, ,
\end{eqnarray}
where  $A=M^2+\omega_p^2-m_e^2$.

For the longitudinal photon (plasmon),  the momentum cutoffs are 
\begin{eqnarray}
|{\bf k}|^\pm &=&\biggl[(2|{\bf P}|^2-2\omega_pE+A)\pm
\sqrt{(2|{\bf P}|^2-2\omega_pE+A)^2
-(4\omega_p^2E^2-4\omega_pEA+A^2)}~\biggr]^{1/2}\, .
\label{eq:ap1limits}
\end{eqnarray}

The use of the leading order dispersion relations yields sufficiently
accurate results except in regions where the plasma frequency 
becomes large so that the ${\bf k}$ integration
enters the region where there are significant deviations from
Eqs. (\ref{eq:tdispersion}) and (\ref{eq:tphotonmass}).

The transverse photon mass in Eq. (\ref{eq:tphotonmass}) is momentum
independent and therefore integration over the photon momentum can
be replaced by integration over energy with limits
provided by Eq. (\ref{Int7}).  We have verified that the two
approaches yield the same result for this case.  However, the
longitudinal mass in Eq. (\ref{eq:lphotonmass}) is momentum dependent
and integrating over momentum is a better choice with the limits
obtained iteratively from Eq. (\ref{eq:p1limits}).

The momentum cut-offs $(k^\pm/\omega_p)_{T,L}$ are shown schematically
in Fig. \ref{fig:approxdisp}.  Depending on the density and
temperature of the plasma, the photo-neutrino rate and emissivity
receive significant contributions from space-like momenta (i.e.,
$|{\bf k}| \geq \omega_p$) in the longitudinal case. (In the
nondegenerate regime, the time-like component is sub-dominant by more
than two orders of magnitude.) This is in contrast to the decay of the
longitudinal plasmon, $\gamma^* \rightarrow \nu + \bar\nu$, which is
kinematically forbidden in space-like regions (i.e., $|{\bf k}| \leq
\omega_p$).

\subsection{Notes For Numerical Integration} 
\label{subsec:monte}

For performing numerical integration, it is convenient to recast
the integrals in Eq. (\ref{Int8}) into dimensionless forms by the
variable transformations 
\begin{eqnarray}\nonumber
E_q = E_\nu^c\,\,x_1,\,\,\quad\,\,
E_{q^\prime} = E_\nu^c\,\,{x_2},\,\,\qquad\,\,
\theta_{qq\prime}& = &\pi\,\,{ x_3}\, , \\\nonumber
E_{p^\prime} =  E^l_{p^\prime}+(E^h_{p^\prime}-E^l_{p^\prime})\,\,{x_4},\,\,\quad\,\,
{\bf |k|} = {\bf |k|}^-+({\bf |k|}^+-{\bf |k|}^-)\,\,{x_7}\, , \\
\phi_e = 2\pi\,\,{x_5},\,\,\quad\,\,
\theta_e =\pi\,\,{ x_6},\,\,\quad\,\,
\phi_k = 2\pi\,\,{ x_8}
\end{eqnarray}
which results in 
\begin{eqnarray}\nonumber
\label{Int12}
Q&=&\frac{(\pi{E_\nu^c})^4}{(2\pi)^8}
\int_{0}^{1} x_1\, dx_1\,\,
\int_{0}^{1} x_2\, dx_2\,\,
\int_{0}^{1} \sin\,\theta_{qq\prime}\,dx_3\,\,
\int_0^1 {\bf |p|}^\prime\, (E^h_{p^\prime}-E^l_{p^\prime})\,\,dx_4
 \int_{0}^{1}\sin\,\theta_e\, dx_6\,\,\\&&\times
\int_{0}^{1}\,dx_5\,\,[1-F_e(E_{p^\prime)}]\,\,
(E_q+E_{q^\prime})\,\,I(p^\prime,q, q^\prime) \\ 
I(p^\prime,q,q^\prime) &=& \frac{1}{8\pi}
\int_0^{1}\frac{{\bf |k|}}{\omega} \,\,
({\bf |k|}^+-{\bf |k|}^-)\,\,{dx_7}\,\,\int_0^{1} dx_8\,\,
F_{\gamma}(\omega)\,\,F_e(E_p)\,\frac{1}{{\bf |P|}}\,\, 
\sum_{s,\epsilon} |\mathcal{M}^{T\,(L)}|^2 \,.
\label{Int42}
\end{eqnarray}
The 8-dimensional integral above is readily integrated by Monte Carlo
methods. Although importance sampling reduces the variance of the
final result, a flat sampling with suitable cut-offs of troublesome
integrands yields equally good results. Such cut-offs can be easily
identified through physical reasoning. 

In the case of degenerate 
electrons, the distribution function $F_e(E_p, T)$ and
the Pauli blocking factor $1-F_e(E_{p^{\prime}}, T)$ ensure that the
relevant integrand peaks when the value of the electron energy is
close to the chemical potential $\mu_e$.  In other words, electrons
primarily play the role of a spectator in this process, originating
chiefly from the vicinity of the Fermi surface and reabsorbed into
nearby available states.  
In the partially degenerate and nondegenerate cases, 
the width of the integrand is
governed by the temperature $T$.  We employ a flat sampling around the
electron chemical potential $\mu_e$ and the width is taken to be
$\simeq 10T$, which gives the range of energy integration for the
outgoing electron to be $E_{p^{\prime}}^\pm=\mu_e \pm 10T$.  Note,
however, that $m_e$ is the natural lower limit for the electron's
energy, and should be employed when necessary.

For neutrinos, the natural upper cutoff is given by the energy that is
available in the medium, which in turn is given by the temperature.
We can also expect that energy will be simply transformed from the
photon to the two neutrinos, since the electron generally plays the
role of a perturbed spectator. The energy released in the form of a
$\nu\bar\nu$ pair cannot greatly exceed the energy scale dictated by
the temperature and setting $E_\nu^c \simeq (10-15)T$ makes a good
choice for the integration cutoff.  This can also confirmed by
inspecting the integrand $dQ/dE_\nu$: the integrand has a maximum
around zero and decays roughly by 2 orders of magnitude for $E_\nu >
10T$.

For certain physical conditions in the plasma, $T\gtrsim m_e$, and for
densities in the range $10^{10} \lsim \rho_B Y_e/{\rm g~ cm^{-3}}
\lsim 10^{12}$, numerical problems are encountered in using a Monte
Carlo procedure to integrate Eq. (\ref{Int8}) (see also
\cite{BEAUDET1,BEAUDET2,DICUS1,BOND1,SCHINDER1,ITOH} in which such
difficulties have been reported).  In Appendix B, we discuss the cause
and remedy of this longstanding problem.

\section{RESULTS}

Our discussion will be restricted to the case of an equilibrium plasma
in which the net negative electric charge of electrons and positrons
is cancelled by a uniform positively charged background of protons,
alpha particles, and heavier ions.  The equation of state and the
phase structure of matter, and the abundances of the various
constituents including those of dripped neutrons at sub-nuclear
densities are determined by the minimization of free energy \cite{LS}.

We will present our results for the total and
differential emissivities as a function of the mass density of protons
in the plasma, $\rho_BY_e = m_pn_e$, where $m_p$ is the proton mass,
$Y_e = n_e/n_B$ is the net electron fraction ($n_B$ is the baryon
number density) , and $n_e$ is net electron number density
\begin{eqnarray}
n_e(T, \mu_e)&=&\frac{1}{\pi^2}\int_0^\infty dp\,\,  p^2 \left(F_{e^-} -
F_{e^+} \right) \,,
\label{MU}
\end{eqnarray}
which is simply the difference between the $e^-$ and $e^+$ number
densities. (The quantity $\rho_BY_e$ is the same as $\rho/\mu_e$ used
in prior works including Ref. \cite{BRAATEN1}.)  Given $n_e$, this
expression can be inverted to determine the chemical potential $\mu_e$
at a given temperature.

\begin{figure}[ht]
\begin{picture}(220, 440)(0, 0)
\put(0, 182){\includegraphics[width=0.45\textwidth]{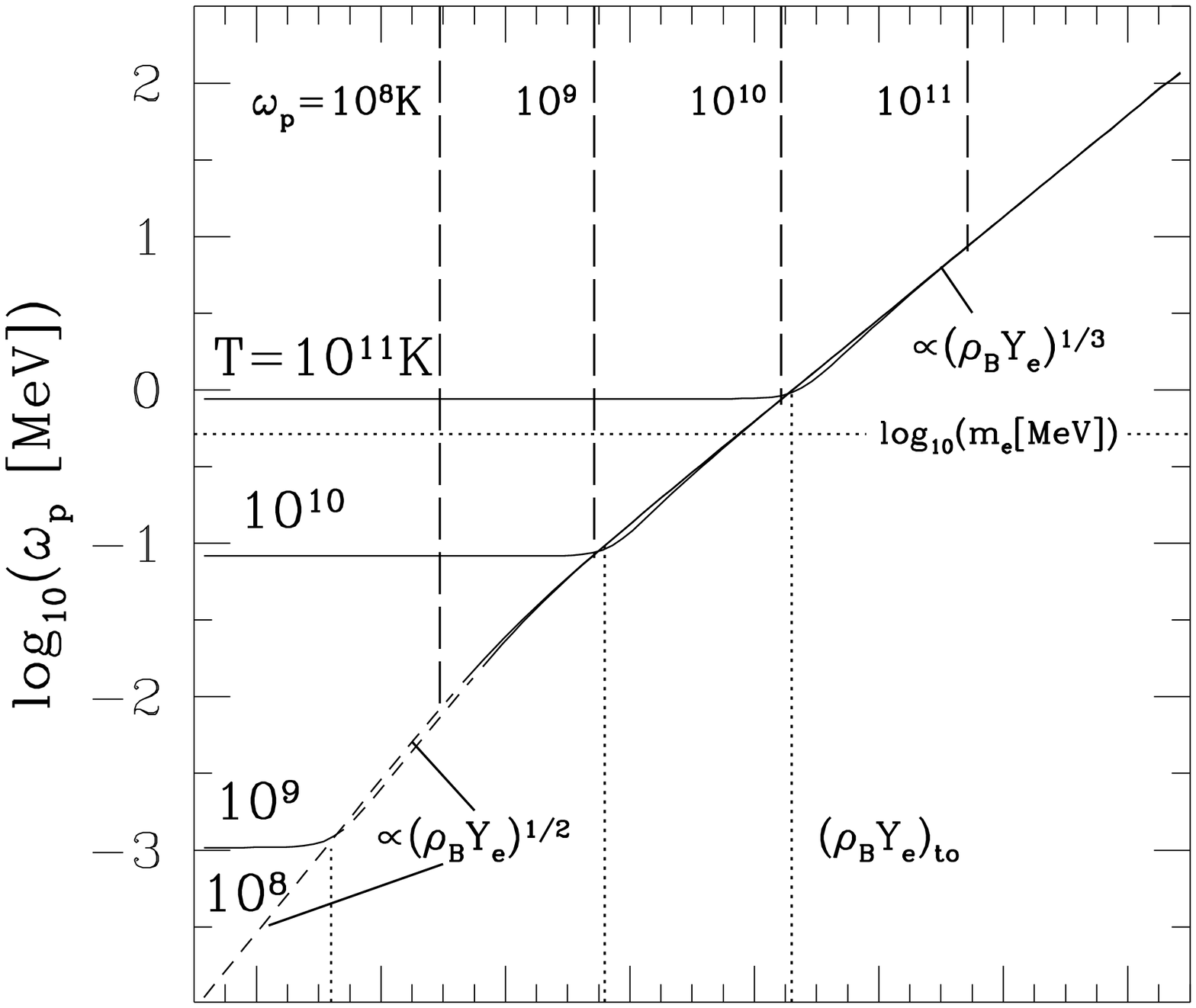}}
\put(180, 282){(a)}
\put(0, 0){\includegraphics[width=0.45\textwidth]{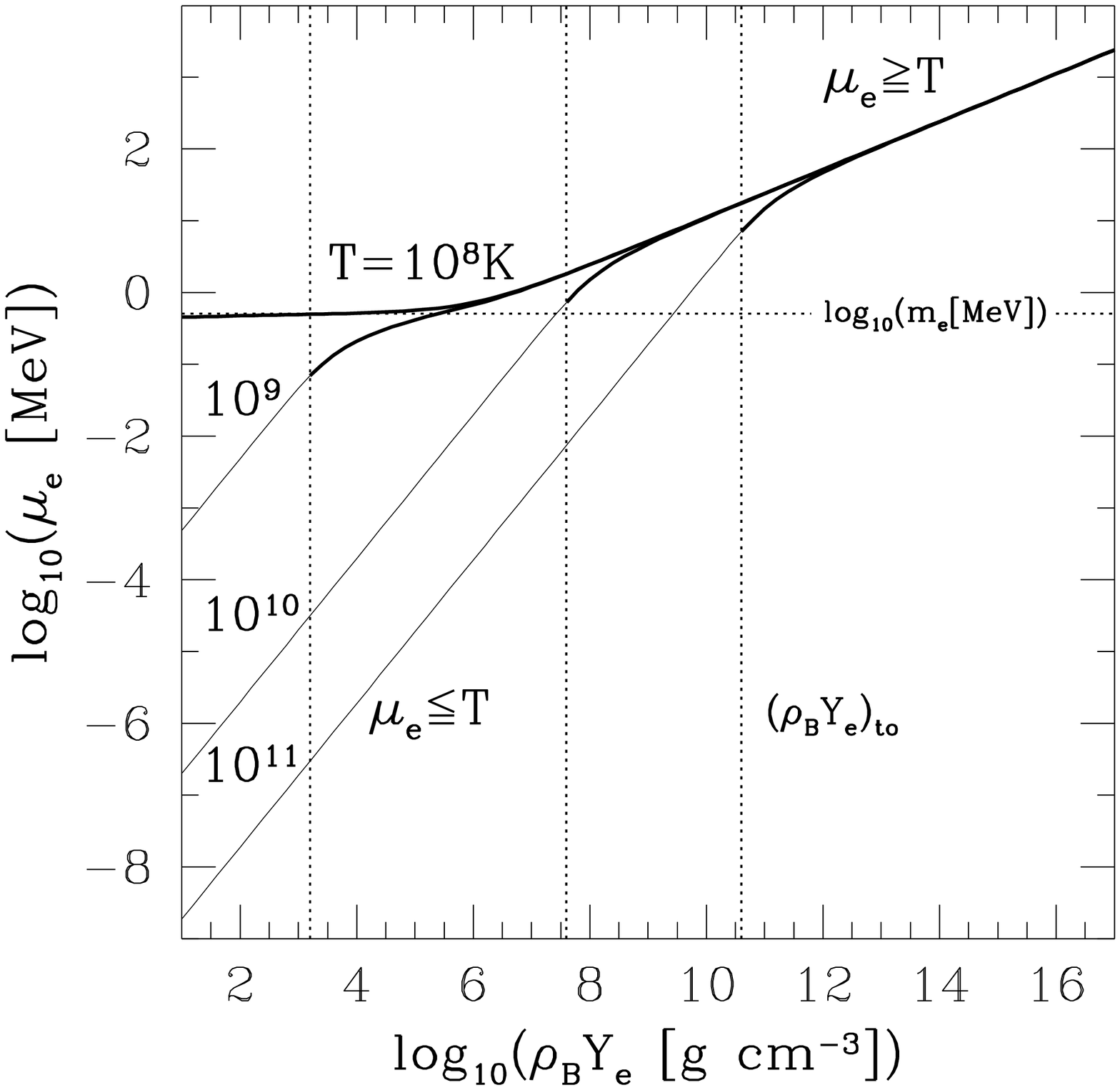}}
\put(180, 120){(b)}
\end{picture}
\caption{The plasma frequency $\omega_p$ (upper panel) and the
chemical potential $\mu_e$ (lower panel) as functions of density for
select temperatures.  The vertical dashed lines in panel (a) indicate
the densities at which $\omega_p = T$.  The vertical dotted lines mark
the turn-on densities at which the plasma frequency and the neutrino
emissivity (Fig.~\ref{fig:qphoto}) abruptly change their behavior from
being independent of density. }
\label{WPMUFIGURE}
\end{figure}

In order to gain a qualitative understanding of the basic features of
$Q_T$ and $Q_L$ in terms of the intrinsic properties of the plasma, it is
instructive to inspect the behaviors of the chemical potential $\mu_e$
and plasma frequency
$\omega_p$ (this is the characteristic energy scale generated by
interactions in the medium) as $T$ and $\rho_BY_e$ are varied.
Figure \ref{WPMUFIGURE} shows $\mu_e$ (lower panel) obtained by inverting
Eq.~({\ref{MU}) and $\omega_p$ (upper panel) calculated from
Eq.~(59) of Ref. \cite{RIP02}. 
The results of the total emissivities are readily
interpreted on the basis of the trends observed in
Fig. \ref{WPMUFIGURE}. 

Noteworthy features of the plasma frequency in panel (a) of 
Fig.~\ref{WPMUFIGURE} are: \\
(1) $\omega_p$ is independent of $\rho_BY_e$ till the turn-on density
$(\rho_BY_e)_{\rm to}$ is reached (this is at the root of why $Q_T$
and $Q_L$ are 
constant for $\rho_BY_e < (\rho_BY_e)_{\rm to}$), and \\
(2) $\omega_p$ shows a power-law increase for $\rho_BY_e >
(\rho_BY_e)_{\rm to}$, the index depending both on the extent to which
the plasma is in the non-degenerate, partially degenerate or
degenerate regime and on whether electrons are relativistic or
nonrelativistic.
 
The bold and light portions of the various curves in panel (b) of
Fig.~\ref{WPMUFIGURE} mark the regions of densities for which $\mu_e
\geq T$ and $\mu_e \leq T)$, respectively.  Inasmuch as $\mu_e \simeq T$
indicates partial degeneracy of the plasma, the bold and light
portions refer to the degenerate and non-degenerate conditions,
respectively.  For reference, the electron mass, which when compared
with $\mu_e$ or $T$ determines the degree of relativity, is marked by
the horizontal dotted line in this figure. The vertical dotted lines
show the respective locations of the turn-on densities.

The results of Monte Carlo integrations of the transverse and
longitudinal total emissivities in Eq.~(\ref{Int12}) are shown as a
function of baryon density in Fig. \ref{fig:qphoto} for select
temperatures. While the emissivities increase rapidly with
temperature, their behavior with density is more intricate.  At a
fixed temperature, the important characteristics to note are: \\ 
\ni
(1) Both $Q_T$ and $Q_L$ are independent of the density $\rho_BY_e$
until a turn-on density $(\rho_BY_e)_{\rm to}$ is reached, \\ 
\ni (2) For densities larger than this turn-on density, $Q_T$ and
$Q_L$ rise rapidly until they attain their peak values at a density
$(\rho_BY_e)_{\rm peak}$, and \\
\ni (3) For $\rho_BY_e \gg
(\rho_BY_e)_{\rm peak}$, the fall-off with density is exponential. \\
A conspicuous feature to note is that with increasing temperature, the
emissivities show a kinky behavior as they begin to approach their
maximum values. 

In Fig.~\ref{fig:qrate}, the transverse and longitudinal rates,
$\Gamma_T$ and $\Gamma_L$, are shown as functions of density and
temperature. The qualitative features of the rates are similar to
those noted above for the emissivities.

The physical origins of the basic trends in the emissivities and rates
are identified in the following section.

\begin{figure}
\includegraphics[width=0.75\textwidth]{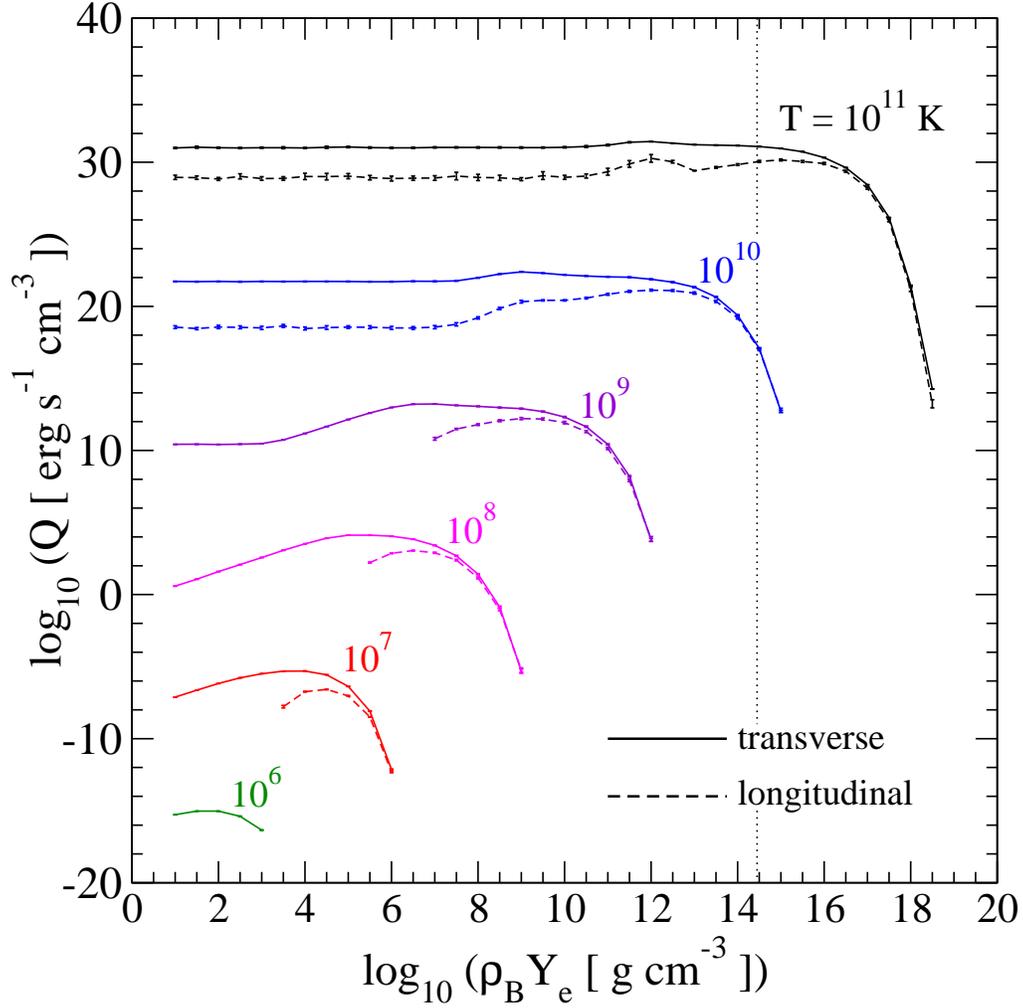}
\caption{Individual contributions from the transverse and longitudinal
channels to the neutrino emissivity as a function of baryon density at
the indicated temperatures. The error bars show the variance of the
Monte Carlo integration of Eq. (\ref{Int12}).  For densities in excess
of nuclear density shown by the dotted vertical line, neutrino production
from strongly interacting particles dominate over QED-plasma
processes. }
\label{fig:qphoto}
\end{figure}

\begin{figure}
\includegraphics[width=0.75\textwidth]{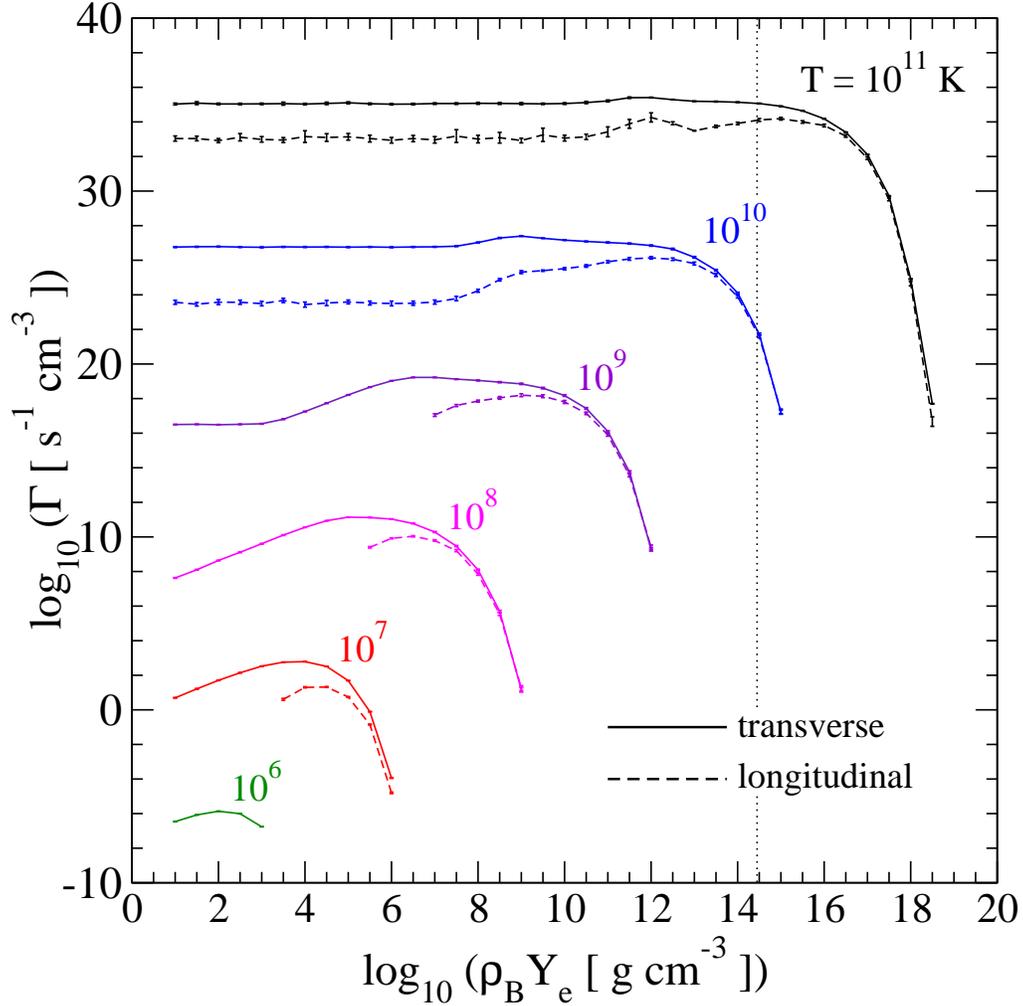}
\caption{Individual contributions from the transverse and longitudinal
channels to the neutrino rates as a function of baryon density at the
indicated temperatures. The error bars show the variance of the Monte
Carlo integration. For densities in excess of nuclear density shown by
the dotted vertical line, neutrino production from strongly interacting
particles dominate over QED-plasma processes. }
\label{fig:qrate}
\end{figure}

\subsection{Analytical Analysis of the Qualitative Behaviours}
\label{sec:approx}
In order to gain a qualitative, and in many cases quantitative,
understanding of the neutrino emissivity as a function of temperature
and density, it is useful to identify the dominant physical scales in
the photoproduction of neutrino pairs under limiting situations.
This can be  achieved in the cases of 
(1) a degenerate plasma, which occurs
for all temperatures at sufficiently high densities, and (2) a
nondegenerate  relativistic plasma in which the density is low, but
the temperature is high.  
Finally, we turn to uncover the origin of the kinks that are
conspicuous at high temperatures, but become less prominent at low
temperatures.

Since our goal here is to provide order of magnitude estimates 
to better understand the exact
numerical results of the preceding section, we will simplify
considerably the integrand in Eq. (\ref{Int8}) by \\
\ni (1) estimating the dominant scales of
energies and momenta of the incoming and outgoing particles, and \\
\ni (2) extensively use physically motivated approximations. \\
These steps are helpful in taming the formidable 8-dimensional
integral in  Eq. (\ref{Int8})
which cannot be otherwise computed analytically.
In the discussion that follows, we will focus on the dominant
transverse channel; the analysis for the longitudinal channel can be
carried along similar lines. 

\subsubsection{The Degenerate Case}
\label{sec:approxa}

For all temperatures at sufficiently high net electron densities
$n_e$, the inequalities 
\begin{eqnarray}
\mu_e \gg T,&\;& \mu_e \gg m_e,\; \nonumber \\ 
\omega_P \gg T,&\;& \omega_P \gg m_e
\end{eqnarray}
are satisfied.  Thus, the electron chemical potential $\mu_e\simeq
{(3\pi^2n_e)}^{1/3}$ and the photon effective mass $\omega_p\simeq
\sqrt{4\alpha/3\pi}\mu_e \simeq \mu_e / 18$ (e.g. see
Refs. \cite{RIP02,BRAATEN1}) are the dominant energy scales
in the problem.  Since $T \ll \mu_e$, the Pauli-blocking of outgoing
electrons ensures that the participating electrons lie close to the
Fermi surface.  In other words, electrons are elastically scattered,
exchanging only the $3$-momentum with the photon and the outgoing
neutrinos.  Under such circumstances, we can expect the electron
energies to be
\begin{equation}
E_p \simeq E_{p^\prime} \simeq {\bf|p|} \simeq {\bf|p^\prime|}\simeq \mu_e
\end{equation}
for both the incoming and outgoing electrons.  This in turn implies
that the photon transfers its entire energy to the $\nu {\bar \nu}$
pair. The photon energy and momentum are
\begin{eqnarray}
\omega\simeq\omega_p \simeq E_q + E_{q^{\prime}} \qquad {\rm and} \qquad 
{\bf|k|}\simeq T \, .
\end{eqnarray}
The characteristic components of the outgoing neutrino 4-momentum
then become $E_q\simeq {\bf|q|} \simeq \omega_p /2$.  These estimates
for the relevant energies and momenta allow us to make an estimate of
the squared matrix element. The generic form of the squared matrix
element can be written as
\begin{eqnarray}
	\sum_{s,\epsilon} {|{\mathcal M}^T|}^2 
	 &\sim& 256 \pi \alpha G_F^2 (C_V^2+C_A^2) 
		\times \frac{1}{(k^2\pm 2 p_i\cdot k)^2}
		\times	\{\rm{a~6^{th}~order~polynomial~in~4-momenta}\}\, ,
\end{eqnarray}
where the dominant terms in the polynomial consist of 2 powers of $k$
and one power each of $q$, $q^{\prime}$, $p$, and $p^{\prime}$. The
contribution from the term involving $m_e^2(C_V^2-C_A^2)$ in the
squared matrix element can be dropped, since $m_e \ll \mu_e$.  By
using power counting combined with our estimates of the relevant
energies and momenta ($q_i \simeq \omega_p/2$, $p_i \simeq \mu_e$, and
$k \simeq \omega_p$), the factor involving the polynomial can be
approximated by $\sim \mu_e^2 \omega_p^4 /4$.

The factor
\begin{eqnarray}
	\frac {1}{\beta_2} &=& \frac {1} {k^2 - 2 p^\prime \cdot k}
\end{eqnarray}
needs more care, since for some ranges of temperature and
density it can develop a resonant behavior (see Appendix B for a
detailed analysis of the physical conditions for which this can occur).  In
the absence of such behavior, its magnitude can be estimated by
averaging this factor over the appropriate angular variable.  
The largest contributions can be determined to be 
\begin{eqnarray}
	\Big\langle \frac{1}{ \beta_1^2} \Big\rangle &\simeq& 
	\Big\langle \frac{1}{\beta_2^2} \Big\rangle \simeq 
	\Big\langle \frac{1}{\beta_1 \beta_2} \Big\rangle 
	\simeq \frac{1}{4 \mu_e^2 \omega_p^2} \, .
\end{eqnarray}	
Using these estimates, the squared transverse 
matrix element can be approximated by
\begin{eqnarray}
	\sum_{s,\epsilon} {| {\mathcal M}^T |}^2  & 
\simeq & 16 \pi \alpha G_F^2 (C_V^2+C_A^2) \omega_p^2 \, .
\end{eqnarray}
The remaining integrals in
Eq. (\ref{Int8}) for the total emissivity 
can be performed along the lines outlined in Sec. \ref{sec:Photo}. 
The transverse emissivity then becomes
\begin{eqnarray}
	Q_T &\simeq& \frac{\alpha G_F^2 (C_V^2+C_A^2)
	\omega_p^2}{4(2\pi)^{9}} \int \frac{d^3q}{E_{q^{\prime}}}
	\frac{d^3q^{\prime}}{E_q} \frac{d^3p^{\prime}}{E_{p^{\prime}}}
	(1-F(E_{p^{\prime}})) \frac{1}{|{\bf P}|} \int_{(E-|{\bf
	P}|)/2}^{(E+|{\bf P}|)/2} dE_p \; F(E_p) \; 
	F_\gamma(E_q+E_{q^{\prime}}) \; (E_q+E_{q^{\prime}})\, .
\label{degQ}
\end{eqnarray}
Since we expect that the most beneficial case for the emissivity is
when $E_p \simeq E_{p^{\prime}} \simeq \mu_e$, we can set $F(E_p)
\rightarrow F(E_{p^{\prime}})$. The $dE_p$ integration simply yields
the range of $E_p$ around $\mu_e$ for which the integrand is significant;
i.e., $\int dE_p\simeq T$. The factor $1/|{\bf P}|=1/|{\bf p}+{\bf
k}|$ can be approximated by $1/\mu_e$, since $|{\bf p}|\simeq \mu_e
\gg |{\bf k}|$. The $dE_{p^{\prime}}$ integrand is sharply peaked
around $E_{p^{\prime}}=\mu_e$, so we can set $E_{p^{\prime}}=\mu_e$
everywhere except in the distribution functions, whence the
$dE_{p^{\prime}}$ integration becomes straightforward. After
introducing new variables $x\equiv E_q/T$ and $y\equiv
E_{q^{\prime}}/T$, and the constant $a\equiv \omega_p/T$, Eq.~(\ref{degQ})
becomes
\begin{eqnarray}
	Q_T &\simeq& \frac{2\alpha G_F^2 (C_V^2+C_A^2)
	\omega_p^2 T^7}{(2\pi)^{6}}
	\iint_0^\infty dxdy\, x y (x+y)\, e^{-(x+y)}\, \theta(x+y-a)\, ,
\end{eqnarray}
where the factor $\theta(x+y-a)$ arises from the condition
$E_q+E_{q^{\prime}}=\omega\ge\omega_p$.  The integral above can be
expressed as
\begin{eqnarray}
	I(a)&=& \iint_0^\infty dxdy\, x y (x+y)\, e^{-(x+y)}\,
	\theta(x+y-a) = \frac{2}{3}\, \Gamma(5, a)\, ,
\end{eqnarray}
where $\Gamma(5, a)$ is  the incomplete gamma function:
\begin{eqnarray}
	\Gamma(n, a)=\int_{a}^\infty dx\, x^{n-1}\, e^{-x} = 
{(a)}^{n-1}\,
	e^{-a}\,\sum_{k=0}^{n-1}{\frac{{(a)}^{-k}(n-1)!}{(n-k-1)!}}\, .
\end{eqnarray}
The rightmost expression above is an asymptotic expansion.
Since in this limit $a=\omega_p/T\gg 1$, we can keep only the leading
order term in the asymptotic expansion to obtain
\begin{eqnarray} \label{eq:qtdeg}
	Q_T&\simeq&\frac{4}{3} \, \frac{\alpha G_F^2
	(C_V^2+C_A^2)}{(2\pi)^{6}}\,
	\, {\omega_p}^6 \, T^3 \, e^{-\omega_p/T}\, .
\end{eqnarray}
Although the emissivity from this expression agrees very well
with our numerical results at high densities, it should be emphasized
that its utility lies  in predicting qualitative trends, 
since the exact numerical
prefactor depends strongly on the approximations 
employed. However, the $Q_T\sim {\omega_p}^6 \, T^3 \,
e^{-\omega_p/T}$ dependence can be employed to predict quantitatively the
density at which the emissivity reaches its maximum value. From
Eq. (\ref{eq:qtdeg}),  we obtain 
\begin{eqnarray}\label{eq:qtdegpeak}
	(\rho_BY_e)_{max}\simeq 9.213 \times 10^{12}\,
	{\Big(\frac{T}{{\rm MeV}}\Big)}^3\, \textrm{g cm}^{-3}\, .
\end{eqnarray}
At high temperatures ($T\gtrsim 10^9$ K) the emissivity has another,
slightly higher maximum in the intermediate region where
($\mu_e>T>\omega_p$). This can significantly broaden the peak and mask
the degenerate maximum. However, we can easily recognize the position
of the peak corresponding to Eq.~(\ref{eq:qtdegpeak}) by noting that a peak
occurs just before the exponential decrease. These positions are
accurately predicted by Eq. (\ref{eq:qtdegpeak}) with inputs for
$\omega_p$ and $\mu_e$ from Fig. \ref{WPMUFIGURE}.
At lower temperatures, or in the longitudinal case in the entire
temperature range, this intermediate maximum is absent and we can
clearly confirm the behaviour predicted by Eq. (\ref{eq:qtdeg}).

Employing the same physical reasoning and
analytical approximations that were used in estimating the total transverse
emissivity, the total transverse rate turns out to be 
\begin{eqnarray}  \label{eq:gtdeg}
	\Gamma_T & \simeq & \frac{4}{3} \, \frac{\alpha G_F^2
	(C_V^2+C_A^2)}{(2 \pi)^6}\, \omega_p^2\, T^6 \: \Gamma(4,
	\omega_p/T) \nonumber \\ & \approx & \frac{4}{3} \,
	\frac{\alpha G_F^2 (C_V^2+C_A^2)}{(2 \pi)^6}\, \omega_p^5\,
	T^3 \: e^{-\omega_p/T}\, .
\end{eqnarray}
%

\subsubsection{The Nondegenerate Case}
\label{sec:approxb}
The nondegenerate situation occurs at sufficiently low densities for
which $\mu_e-m_e \ll T$.  In this case, both $Q_T$ and $Q_L$ exhibit a
plateau for temperatures $T \ge 10^9$ K. This behavior of the
emissivities is intimately connected with a similar behavior of the
plasma frequency $\omega_p$ for the corresponding temperatures (see
Fig. \ref{WPMUFIGURE}). In this case, we can expect that a significant
fraction of the electrons participate in energy
exchange.


For $T\gtrsim 10^{10}$ K, we can neglect the electron mass
($m_e=5.93\times10^9$ K) in comparison to $\mu_e$. In the relativistic
regime, net electron density and the plasma frequency are
given by \cite{RIP02,BRAATEN1}
\begin{eqnarray}\label{eq:relwpmu}
n_e &=& \frac{\mu_e}{3\pi^2}\biggl(
\mu_e^2+T^2\pi^2\biggr) \simeq \frac{\mu_e T^2}{3} \nonumber \\
\omega_p^2 &=& \frac{4\alpha}{3\pi}\biggl(\mu_e^2+
\frac{\pi^2T^2}{3}\biggr)\simeq \frac{4\pi \alpha T^2}{9} \, ,
\end{eqnarray}
where the rightmost expressions are valid for $T\gg \mu_e$. In this
case, the characteristic electron energy and momentum are of order
$E_p \simeq {\bf |p|} \simeq T$, and the photon energy and momentum
are $\omega\simeq {\bf |k|} \simeq T$.

Since the temperature becomes the dominant energy scale governing the
neutrino-pair emission process, we can use dimensional analysis to
establish that the neutrino emissivity and rate will scale with
temperature as $T^9$ and $T^8$, respectively.  An analytical
estimation of the numerical prefactors, however, needs some work.  In
this case, we have identified the leading term in the squared
transverse matrix element to be
\begin{eqnarray}
 \sum_{s,\epsilon} {|{\mathcal M}^T|}^2 &\simeq& 256\, \pi\, \alpha\, G_F^2 \,
  (C_V^2+C_A^2) \, 
  \frac{2(p^\prime\cdot k)\big( (p\cdot q)(k \cdot q^\prime) + 
(p\cdot q^\prime)(k \cdot q)\big) } {{(2 p^\prime \cdot k - k^2)}^2}\, .
\end{eqnarray}
The predominance of this term over all other terms is chiefly due to
the resonant nature of the denominator.  The angular dependences involving the
outgoing neutrinos do not have a large effect, whence, to very good
approximation we can write
\begin{eqnarray}
\sum_{s,\epsilon}  {|{\mathcal M}^T|}^2 &\simeq& 256\, \pi\, \alpha\, G_F^2 \,
  (C_V^2+C_A^2) \, \omega E_p E_q E_{q^\prime}
  \frac{(p^\prime\cdot k)}{{(2 p^\prime \cdot k - k^2)}^2}\, .
\end{eqnarray}
We can further use the average value of 
\begin{eqnarray}
  \frac{(p^\prime\cdot k)}{{(2 p^\prime \cdot k - k^2)}^2}
  &\simeq&
  \frac{1}{2}\int_{-1}^1 \frac{E_{p^\prime}\omega - 
|{\bf {p^\prime}}||{\bf k}| x}
  {{(2 E_{p^\prime}\omega - 2 |{\bf p^\prime}||{\bf k}| x -
  \omega_p^2)}^2} 
\, dx\, ,
\end{eqnarray}
where $x$ denotes the cosine of the angle between ${\bf p^\prime}$ and
${\bf k}$. The value of this integral does not depend sensitively on
$\omega_p^2$, since it is negligibly small compared to the other terms
in the denominator. We can therefore set it to zero without
significantly changing the final result.  In the relativistic regime,
we can also ignore the electron mass compared to its typical momentum.
We then find
\begin{eqnarray}
  \frac{(p^\prime\cdot k)}{{(2 p^\prime \cdot k - k^2)}^2}
  &\simeq&
  \frac{1}{2} \frac{1}{4 E_{p^\prime}\omega} 
  \ln\left( \frac{4\omega^2}{\omega_p^2}\right)
  \simeq\frac{1}{E_{p^\prime}\omega}\, ,
\end{eqnarray}
where in obtaining the rightmost result, we have utilized the average
energy of an equilibrated nearly massless photon $\langle \omega
\rangle= 2.7T$ and $\omega_p\simeq T/10$ as appropriate for this
regime.  The working expression for the squared matrix element then
becomes
\begin{eqnarray}
 \sum_{s,\epsilon} {|{\mathcal M}^T|}^2 &\simeq& 256\, \pi\, \alpha\, G_F^2 \,
  (C_V^2+C_A^2) \, \frac{E_p E_q E_{q^\prime}}{E_{p^\prime}}\, .
\end{eqnarray}
The identity
\begin{eqnarray}
  \int \frac{d^3q}{2E_q}\,
  \frac{d^3q^{\prime}}{2E_{q^{\prime}}}\,
  \delta^4(q+q^{\prime}-q_t) \, {E_q\, E_{q^\prime}}
  &=&
  \frac{\pi}{24}\, \Theta(q_t^2)\, (3{q_t^0}^2-|{\bf q_t}|^2)\, ,
\end{eqnarray}
where the four-vector $q_t=(q_t^0,{\bf q_t}) \equiv q+ q^\prime = p + k
-p^\prime$, helps us to write the 
emissivity and the rate as 
\begin{eqnarray}\label{eq:qgammatheta}
  \genfrac{(}{)}{0pt}{}{Q_T}{\Gamma_T}&=&
  \frac{32 \pi^2 \alpha G_F^2 (C_V^2+C_A^2)}{3(2\pi)^{11}}
  \int \frac{d^3p}{2E_p}\,n_F(E_p)\,
  \frac{d^3p^{\prime}}{2E_{p^{\prime}}}\,
  \left(1-n_F(E_{p^{\prime}})\right)\,
  \frac{d^3k}{2\omega}\,n_B(\omega)\, \nonumber \\
  && \times \, \Theta(q_t^2) \, (3{q_t^0}^2-|{\bf q_t}|^2)\,
 \frac{E_p E_q E_{q^\prime}}{E_{p^\prime}}\,
  \genfrac{(}{)}{0pt}{}{E_p+\omega-E_{p^\prime}}{1}\, .
\end{eqnarray}
The remaining integrations are, however, complicated because of the
complex integration boundaries arising from the $\Theta(q_t^2)$
factor. In order to simplify this restriction and the integration over
the $(3{q_t^0}^2-|{\bf q_t}|^2)$ factor, we use 
$\Theta(q_t^2)=\Theta(q_t^0-|{\bf q_t}|)$, and replace
$|{\bf q_t}|$ and $|{\bf q_t}|^2 $ by their average values:
\begin{eqnarray}
\langle |{\bf q_t}| \rangle &=&
  \frac{1}{2} \int_{-1}^1 dy\, 
  \sqrt{(E_q^2 + E_{q^\prime}^2 +2 E_q E_{q^\prime}y)}
  =E_>+\frac{1}{3}\frac{E_<^2}{E_>} \simeq q_t^0/2 \label{eq:ravg}\\
  \langle |{\bf q_t}|^2 \rangle &=&
  \frac{1}{2} \int_{-1}^1 dy\, 
  (E_q^2 + E_{q^\prime}^2 +2 E_q E_{q^\prime}y)
  ={q_t^0}^2-2E_q E_{q^\prime} \simeq {q_t^0}^2/2\label{eq:r2avg}\, ,
\end{eqnarray}
where $y$ denotes the cosine of the angle between the two outgoing
neutrinos.  In Eq. (\ref{eq:ravg}), $E_>$ and $E_<$ are the larger and
smaller of the two energies $E_q$ and $E_{q^\prime}$, respectively. In
the last steps of Eqs. (\ref{eq:ravg}) and (\ref{eq:r2avg}), we have
replaced the neutrino energies with their average value $q_t^0/2$. As a
result, the following simplifications can be made:
\begin{eqnarray}
  \Theta(q_t^2) & \rightarrow & \Theta(q_t^0) \nonumber \\
  3{q_t^0}^2-|{\bf q_t}|^2 & \rightarrow & \langle  3{q_t^0}^2-|{\bf
  q_t}|^2  \rangle = 5{q_t^0}^2/2\, .
\end{eqnarray}
We note that the condition $\Theta(q_t^0)=\Theta(E_p+w-E_{p^\prime})$ has
the physical interpretation that the outgoing electron energy cannot
exceed the total incoming energy.

The emissivity and rate can now be expressed in terms of the simple expressions
\begin{eqnarray}
  \genfrac{(}{)}{0pt}{}{Q_T}{\Gamma_T}&=&
  \frac{20 \alpha G_F^2 (C_V^2+C_A^2)}{3(2\pi)^{6}}
  \, \genfrac{(}{)}{0pt}{}{T^9{\mathcal I}_Q }{T^8{\mathcal I}_\Gamma }\, ,
\end{eqnarray}
where the dimensionless constants are
\begin{eqnarray}
  \genfrac{(}{)}{0pt}{}{{\mathcal I}_Q}{{\mathcal I}_\Gamma}
  &=&
  \int_0^\infty \frac{dx \, x^2}{e^x+1} 
  \int_0^\infty \frac{dy \, y}{e^y-1} 
  \int_0^{x+y} \frac{dz}{e^{-z}+1}
  \genfrac{(}{)}{0pt}{}{(x+y-z)^3}{(x+y-z)^2}  \, ,
\end{eqnarray}
and the integration variables are $x \equiv E_p/T$, $y \equiv w/T$, and $z
\equiv E_{p^\prime}/T$. 
A numerical evaluation of these integrals yields
\begin{eqnarray}
\label{iqig} 
 \genfrac{(}{)}{0pt}{}{{\mathcal I}_Q}{{\mathcal I}_\Gamma}
  &=&    \genfrac{(}{)}{0pt}{}{775.54}{136.50}  
\end{eqnarray}
and the ratio ${\mathcal I}_Q/{\mathcal I}_T = 5.68$. 
Analytical approximations to these integrals can be obtained 
by the replacement
\begin{eqnarray}
\frac{1}{e^{-z}+1} &\rightarrow& 1
\end{eqnarray}
with the result 
\begin{eqnarray}
{{\mathcal I}_Q}&=& \frac{63}{256}\Gamma(7)\zeta(7)\Gamma(2)\zeta(2)
	+\frac{37}{32}\Gamma(6)\zeta(6)\Gamma(3)\zeta(3)
	+\frac{73}{32}\Gamma(5)\zeta(5)\Gamma(4)\zeta(4)
	= 1002.1 \\
{{\mathcal I}_\Gamma}&=&\frac{31}{96}\Gamma(6)\zeta(6)\Gamma(2)\zeta(2)
	+\frac{19}{16}\Gamma(5)\zeta(5)\Gamma(3)\zeta(3)
	+\frac{7}{8}\Gamma^2(4)\zeta^2(4)
	= 172.74 
\end{eqnarray}
($\Gamma(n)$ and $\zeta(n)$ are the Gamma and Riemann's zeta
functions, respectively) and the ratio ${\mathcal I}_Q/{\mathcal I}_T
= 5.8$ which are reasonably close to the exact numerical results.
Using the results in Eq.~({\ref{iqig}), the emissivity and the rate in
the nondegenerate case are
\begin{eqnarray}
\label{qtnd}
  Q_T &\simeq& 4.35\times 10^{22} 
  {\left(\frac{T}{\rm MeV}\right)}^9 \, 
  \rm{erg}\, \rm{cm}^{-3}\, \rm{s}^{-1}\,,\\
  \Gamma_T &\simeq& 4.78\times 10^{27} 
  {\left(\frac{T}{\rm MeV}\right)}^8 \, 
  \rm{cm}^{-3}\, \rm{s}^{-1}\, .
\label{gtnd}
\end{eqnarray}
The above estimate for the emissivity is in good agreement with the exact
numerical results presented in Fig.  \ref{fig:qphoto}.

As the density increases, the electron chemical potential and the
plasma frequency both begin to increase, and become strongly density
dependent. As a result, the emissivity also acquires a strong 
density dependence. We can expect this change of behaviour to occur
when the term involving the chemical potential in
Eq. (\ref{eq:relwpmu}) becomes comparable to the term involving
$T^2$. In this case,
\begin{eqnarray}
{\displaystyle (n_e)_{\rm to} }& \simeq &{\displaystyle  \frac{T^3}{2.72} } 
\quad {\rm or}  \\  
{\displaystyle \left({\rho_B}{Y_e}\right)_{\rm to} }
& \simeq & {\displaystyle8 \times 10^7 \left(\frac{T}{{\rm MeV}}\right)^3}~
{\rm g~cm^{-3}} \nonumber \\
	\Rightarrow \log_{10}({\rho_B}{Y_e})
		&=& \left\{
		\begin{array}{ll}
			10.7 & \text{at $T=8.6$ MeV ($10^{11}$ K)} \\
			7.7 & \text{at $T=0.86$ MeV}
		\end{array} \right. \nonumber
\end{eqnarray}
which agrees closely with the turn-on densities in
Fig. \ref{fig:qphoto}.

\subsubsection{Intermediate Case}
\label{sec:approxc}
In the case that $\mu_e>T>\omega_p$, the emissivity $Q_T$ acquires a strong
density dependence.  In addition, a conspicuous maximum occurs prior
to entering the strongly degenerate regime.
In this density range, the dominant energies are 
\begin{eqnarray}
  \text{electrons: } && E_p \simeq {\bf|p|} \simeq \mu_e \nonumber \\
  \text{photon: } && \omega \simeq {\bf|k|} \simeq T \nonumber \\
  \text{neutrinos:} && E_q \simeq {\bf|q|}  \simeq T\, .
\end{eqnarray}
With these energy scales, the squared matrix element can be
estimated to be 
\begin{eqnarray}
  	\sum_{s,\epsilon} 
{|{\mathcal M}^T|}^2  &\sim& 256 \pi \alpha G_F^2 (C_V^2+C_A^2) \,
	\frac{1}{4\mu_e T \omega_p^2}\, T^4\, \mu_e^2\, .
\end{eqnarray}
The phase space integration can be performed similarly to the
degenerate case with the result
\begin{eqnarray}\label{eq:intqtphoto1}
	Q_T & \simeq & \frac{32}{3}\,
		\frac{\alpha G_F^2 (C_V^2+C_A^2)}{(2\pi)^{6}}\,
		\frac{T^9 \mu_e^2}{\omega_p^2}\,
		\zeta(5)\, \Gamma(5, \omega_p/T)\, .
\end{eqnarray}
In the region of maximum emissivity, the ratio $\omega_p/T\ll 1$ for
the two highest temperatures ($10^{10}$ and $10^{11}$ K).  This allows
us to expand the incomplete gamma function in terms of this small
parameter so that
\begin{eqnarray}\label{eq:intqtphoto2}
	Q_T & \simeq & \frac{768}{3}\,
		\frac{\alpha G_F^2 (C_V^2+C_A^2)}{(2\pi)^{6}}\,
		\frac{T^9 \mu_e^2}{\omega_p^2}\,
		e^{-\omega_p/T} .
\end{eqnarray}
For $T=8.6$ MeV, the maximum occurs around $\rho_B Y_e=10^{12}$ g
cm$^{-3}$, where we have used $\mu_e=46.8$ MeV and $\omega_p=2.74$
MeV. By combining these values with Eq. (\ref{eq:intqtphoto2}), the
emissivity can be found to be $\log_{10}Q_T=32.06$ in very good
agreement that in Fig. \ref{fig:qphoto}.

For $T=0.86$ MeV, $Q_T$ exhibits a maximum around $\rho_B Y_e=10^{9}$
g cm$^{-3}$. In this case, $\mu_e=4.71$ MeV and $\omega_p=0.274$
MeV. These values yield $\log_{10}Q_T=23.0$, which is slightly larger
than the exact numerical result.

The origin of the secondary peak lies in the resonant character of the
factor 
\begin{equation}
\frac{1}{\beta_2^2} = 
\frac {1}{(2E_p \omega-2{\bf p}\cdot{\bf k}-\omega_p^2)^2} 
\end{equation}
for the case in which the plasma frequency
becomes negligible, but $E_p\simeq {\bf |p|}$ and $\omega\simeq {\bf |k|}$. 
Such conditions cannot be satisfied in a strongly degenerate medium in
which $\omega\simeq\omega_p \gg k\simeq T$ or at low temperatures for
which $E_p \simeq m_e$ and $|{\bf p}|  \simeq T \ll m_e$. Hence, as the
temperature decreases this enhancement becomes reduced and at
$T\lesssim 10^8$ K it is completely absent.

This resonant structure also gives rise to numerical problems
at high temperatures and is further discussed in Appendix
\ref{sec:singu}. Unless suitably accounted for, 
this factor enhances the variance of the Monte Carlo integration by
large factors.

\subsection{Typical Neutrino Energies}

The mean neutrino plus anti-neutrino energy can be characterized by 
the ratio of the total emissivity to the total rate. In this section, we 
analyze 
\begin{eqnarray}
\label{eq:eavg}
	{\langle E_{\nu \bar \nu}\rangle}_T 
	& = & \frac{Q_T}{\Gamma_T}\, 
\end{eqnarray}
for the transverse case. The analysis
for the longitudinal case can be performed in a similar fashion.
Utilizing the numerical results from Eq.~(\ref{Int12}) and its counterpart
for the rate, the exact results for $\langle E_{\nu \bar
\nu}\rangle_T$ versus density at various temperatures are shown by
the solid lines in Fig. \ref{fig:eavg}.  The discussion below is aimed
toward a qualitative understanding of the basic trends in this
figure in limiting cases.

In the degenerate case, the results in Eqs.~(\ref{eq:qtdeg}) and 
(\ref{eq:gtdeg})
set the average neutrino pair energy to be
\begin{eqnarray}
\label{eq:enudeg}
	{\langle E_{\nu \bar \nu}\rangle}_T 
	& \simeq & \omega_p \, 
\end{eqnarray}
in line with the assumption that $E_q+E_{q^{'}} \simeq \omega_p$ used
in the approximation procedure.
For the two highest temperatures $T/{\rm MeV}=0.86$ and 8.6 shown in   
Fig. \ref{fig:eavg}, $\langle E_{\nu\bar\nu} \rangle =6.8T$ and $7T$,
respectively. 
From the results in Eqs. (\ref{qtnd}) and (\ref{gtnd}) 
in Sec. \ref{sec:approxb}, we obtain
\begin{eqnarray}\label{eq:enurel}
	{\langle E_{\nu \bar \nu}\rangle}_T & \simeq & 5.8\, T\,
\end{eqnarray}
in the non-degenerate relativistic case, which, considering  the
approximations made, is a reasonable estimate.

It is instructive to compare
the mean neutrino pair energy ${\langle E_{\nu \bar \nu}\rangle}_T$
with the characteristic energy of the in-medium photons in the plasma.
The average photon energy in the plasma is given by
\begin{eqnarray}
\label{eq:Bavg}
\langle\omega_T\rangle &=& 
\frac{\displaystyle\int_0^\infty dk~k^2 \omega_T \, F_{\gamma}(\omega_T,T)}
{\displaystyle\int_0^\infty dk \, k^2 F_{\gamma}(\omega_T,T)} 
\simeq \; 
 \frac{\omega_p^2}{8 T} \;
\frac { {\displaystyle\sum_{j=1}^\infty }{\displaystyle\biggl[}K_4(jy) -
K_0(jy){\displaystyle\biggr]} } 
{ {\displaystyle\sum_{j=1}^\infty }K_2(jy)/j}~ \,,  
\end{eqnarray}
where the rightmost relation is obtained upon setting $\omega_T^2 -
k^2 \simeq m_T^2 \approx \omega_p^2$ (see also Ref. \cite{RIP02}).  Simpler
results ensue for the extreme relativistic and nonrelativistic cases:
\begin{eqnarray}
\label{eq:Brel}
\langle \omega_T \rangle &\simeq& 2.7~T \qquad \quad {\rm for} \quad T \gg
\omega_p  \\ 
&\simeq& \omega_p + \frac 32 ~T  \quad \, {\rm for} \quad T \ll \omega_p \,.
\end{eqnarray}
The dashed curves in Fig. \ref{fig:eavg} show expectations based on
Eq. (\ref{eq:Bavg}).

Comparing the result in Eq. (\ref{eq:enurel}) with that in
Eq. (\ref{eq:enudeg}), we confirm that in the degenerate case the process
proceeds as the decay of a massive photon into two neutrinos, with
electrons and positrons emerging with modified angles in their final
states.  The numerical results in the figure also support this
expectation.  Notice, however, that as the density is progressively
increased, the average neutrino energy becomes somewhat smaller than
the average photon energy.  This can be attributed to the fact that
some of the available energy is taken by the outgoing electrons.

In the relativistic case, the average neutrino pair energy
(Eq. (\ref{eq:enurel})) is significantly enhanced relative to the
average photon energy (Eq. (\ref{eq:Brel})). This can be attributed to
the different energy weightings of the squared matrix element in $Q_T$
and $\Gamma_T$ not accounted for in Eq. (\ref{eq:Bavg}) and to the
fact that electrons and positrons impart some of their energy to
the neutrino pair.

\begin{figure}[h!t]
\begin{center}
\includegraphics[width = 0.5\textwidth]{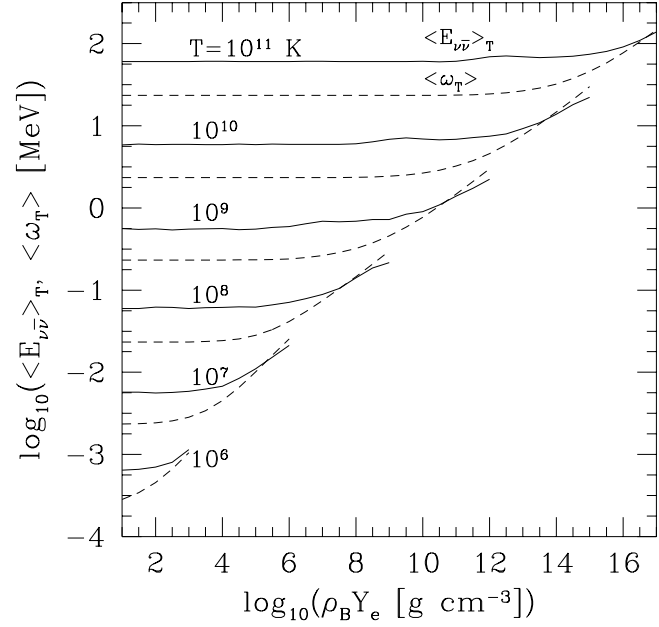}
\end{center}
\caption{Average transverse photon and neutrino pair energies as
 functions of density and temperature.}
\label{fig:eavg}
\end{figure}

\section{ Kernels for Neutrino Transport Calculations}
\label{sec:PKER}
The matrix elements derived in Sec. \ref{sec:Photo} enable us to obtain 
the source and sink terms 
associated with the photo-neutrino process which
are required in neutrino transport calculations.
The evolution of the neutrino distribution function $f$, generally
described by the Boltzmann transport equation in conjunction with
hydrodynamical equations of motion together with baryon and lepton
number conservation equations is 
\begin{eqnarray}
\label{BOLTZMANN}
\frac{\partial f}{\partial t}+v^i\frac{\partial f}{\partial x^i}+
\frac{\partial (f F^i)}{\partial p^i}
=B_{EA}(f)+B_{NES}(f)+B_{\nu{\cal N}}(f)+B_{TP}(f) \,.
\end{eqnarray}
Here, $F^i$ is the force acting on the particle  and we have 
ignored general  relativistic effects for
simplicity (see, for example, Ref. \cite {BRUENN1} for full details).
The right hand side of the above equation is the neutrino source term in which, 
$B_{EA}(f)$ incorporates neutrino emission and
absorption processes, $B_{NES}(f)$ accounts for the neutrino-electron
scattering process, $B_{\nu{\cal N}}(f)$ includes scattering of
neutrinos off nucleons and nuclei, and $B_{TP}(f)$ considers
the thermal production and absorption of neutrino-antineutrino pairs.

In this section, we obtain the contribution from the photo-neutrino process
to the thermal production term $B_{TP}$. 
In doing so, we follow closely Ref. \cite{RIP02} where kernels 
for the plasma process  and Ref. \cite{BRUENN1} in which 
neutrino pair production from $e^+e^-$ annihilation were obtained.
Suppressing the dependencies on $(r,t)$ for notational simplicity, 
the source term for the photo-neutrino process can be written as
\begin{widetext}
\begin{eqnarray}\label{BTP}\nonumber
B\big(f(\mu_q,E_q)\big) &=& 
\biggl[1-f(\mu_q,E_q)\biggr]\frac{1}{(2\pi)^3}
\int_0^\infty E^2_{q^\prime} \,\,dE_{q^\prime} \int_{-1}^1 
d\mu_{q^\prime} \int_0^{2\pi}d\phi_{q^\prime} \,R^p
\biggl(E_q,E_{q^\prime},\cos\,\theta_{qq\prime}\biggr)
\,\biggl[1-\bar{f}(\mu_{q^\prime},E_{q^\prime})\biggr]\\
&-&  f(\mu_q,E_q)\frac{1}{(2\pi)^3}
\int_0^\infty E^2_{q^\prime}  \,\,dE_{q^\prime} \int_{-1}^1 d\mu_{q^\prime} 
\int_0^{2\pi}d\phi_{q^\prime}
\,R^a\biggl(E_q,E_{q^\prime},\cos\,\theta_{qq\prime}\biggr)
\,\bar{f}(\mu_{q^\prime},E_{q^\prime}) \,,
\end{eqnarray}
\end{widetext}
where the first and the second terms correspond to the source
(neutrino gain) and sink (neutrino loss) terms, respectively. Angular
variables $\mu_{q(q^\prime)}\equiv \cos\theta_{q(q^\prime)}$ and 
$\phi_{q(q^\prime)}$ are defined with
respect to the $z$-axis that is locally set parallel to the outgoing
radial vector ${\bf r}$. The angle $\theta_{qq\prime}$ between the
neutrino and antineutrino pair is related to $\theta_q$ and $\theta_{q^\prime}$
through 
\begin{eqnarray}
\cos\theta_{qq\prime} &=&
\mu_q\mu_{q^\prime}+\sqrt{(1-\mu_q^2)(1-\mu_{q^\prime}^2)}~\cos(\phi_q-\phi_{q^\prime}) \,.
\end{eqnarray}
The production kernel is given by
\begin{equation}
R^{p}(E_q, E_{q^\prime}, \cos\theta_{qq\prime})=
\int\frac{2d^3{\bf p}}{(2\pi)^3} \frac {F_e(E_p)}{2E_p} 
\int\frac{\xi d^3{\bf k}}{(2\pi)^3} \frac {F_{\gamma}(\omega)}{2\omega} 
\int\frac{d^3{\bf p^{\prime}}}{(2\pi)^3}\frac{ [1-F_e(E_{p^{\prime}})]}{2E_{p^{\prime}}}
\frac{(2\pi)^4}{4 E_q E_{q^{\prime}}} \delta^4(p+k-p^{\prime}-q-q^{\prime})
\frac{1}{\zeta}\sum_{s,\epsilon} |{\mathcal{M}}|^2\,.
\label{kernel}
\end{equation}
The corresponding expression for the absorption kernel $R^{a}(E_q,
E_{q^\prime}, \cos\theta_{qq\prime})$ can be obtained by the
replacements
\begin{eqnarray}
\nonumber
F_e(E_p) \rightarrow 1-F_e(E_p)\,, \qquad
F_{\gamma}(\omega) \rightarrow 1+F_{\gamma}(\omega)\,, \qquad
{\rm and} \qquad
1-F_e(E_{p^{\prime}})\rightarrow F_e(E_{p^{\prime}})\, .
\label{Constant}
\end{eqnarray}
The angular dependences in the kernels $
R^{\genfrac{}{}{0pt}{}{p}{a}}(E_q, E_{q^\prime}, \cos\theta_{qq\prime})$
are often expressed in terms of Legendre polynomials as
\begin{eqnarray}\label{RLegendre}
R^{\genfrac{}{}{0pt}{}{p}{a}}(E_q, E_{q^\prime}, \cos\theta_{qq\prime})&=&
\sum_{l=0}^{\infty} \frac{2l+1}{2} \Phi^{\genfrac{}{}{0pt}{}{p}{a}}_l(E_q, E_{q^\prime}) 
P_l (\cos\theta_{qq\prime}) \,, 
\end{eqnarray}
where the Legendre coefficients
$\Phi^{\genfrac{}{}{0pt}{}{p}{a}}_l(E_q, E_{q^\prime})$ depend exclusively on
energies.
 
From Eq. (\ref{kernel}), it is evident that the kernels are related to
the neutrino rates and emissivities.  We first consider the production
kernel $R^{p}(E_q, E_{q^\prime}, \cos\theta_{qq\prime})$.  The
corresponding analysis for the absorption kernel $R^{a}(E_q,
E_{q^\prime}, \cos\theta_{qq\prime})$ can be made along the same
lines.  The neutrino production rate is given by
\begin{eqnarray}\label{rate}\nonumber
\Gamma&=&
\int\frac{2\,d^3{\bf p}}{(2\pi)^3} \frac {F_e(E_p)}{2E_p} 
\int\frac{\xi\,d^3{\bf k}}{(2\pi)^3} \frac {F_{\gamma}(\omega)}{2\omega} 
\int\frac{d^3{\bf p^{\prime}}}{(2\pi)^3} \frac{ [1-F_e(E_{p^{\prime}})]}{2E_{p^{\prime}}}
\int\frac{d^3{\bf q}}{(2\pi)^3} \frac{1}{2E_q}
\int\frac{d^3{\bf q^{\prime}}}{(2\pi)^3}  \frac{1}{2E_{q^{\prime}}} 
\\ &&\times
\,\, (2\pi)^4 \delta^4(p+k-p^{\prime}-q-q^{\prime})\,\frac{1}{\zeta}
\sum_{s,\epsilon} |{\mathcal{M}}|^2\,.
\\&=& \int \frac{d^3 {\bf q}}{(2\pi)^3} \frac{d^3{\bf q}^{\prime}}{(2\pi)^3} 
R^{p}(E_q, E_{q^\prime}, \cos\theta_{qq\prime})\,, 
\end{eqnarray}
which defines the kernel $R^p(E_q, E_{q^\prime}, \cos\theta_{qq\prime})$ and is to be
identified with that in Eq.~(\ref{kernel}).
The emissivity $Q$ can also be cast in terms of $R^p$ using 
\begin{equation}
Q=\int \frac{d^3{\bf q}}{(2\pi)^3} \frac{d^3 {\bf q}^{\prime}}{(2\pi)^3} (E_q+E_{q^\prime}) 
R^{p}(E_q, E_{q^\prime}, \cos\theta_{qq\prime})\,. \label{Remiss}
\end{equation}
Equations (\ref{rate}) and (\ref{Remiss}) can be inverted to obtain
\begin{eqnarray}\label{RQ}
R^{p}(E_q, E_{q^\prime}, \cos\theta_{qq\prime}) & = &
\frac{8\pi^4}{E_q^2 E_{q^\prime}^2}~\frac{d^3 \Gamma}
{dE_q dE_{q^\prime} d\cos\theta_{qq\prime}}\nonumber \\
&=&\frac{8\pi^4}{E_q^2 E_{q^\prime}^2 (E_q+E_{q^\prime})}~\frac{d^3 Q}{dE_q dE_{q^\prime} d\cos\theta_{qq\prime}}\,.
\end{eqnarray}
Combining Eq. (\ref{dInt1}) with the above equation,  
the production kernel can be written as 
\begin{equation}
R^{p}(E_q, E_{q^\prime}, \cos\theta_{qq\prime})= 
\frac{\pi^3}{(2\pi)^6 E_q E_{q^\prime}}
\int_0^\infty \frac{{\bf |p^\prime|}^{2}}
{E_{p^\prime}}d{\bf |p|}^\prime \int_{-1}^{1}
 d\cos\,\theta_{e}\int_0^{2\pi} d\phi_{e}[1-F_e(E_{p^\prime})]
I(p^\prime,q, q^\prime)
\label{kerfinal}
\end{equation}
where $I(p^\prime,q,q^\prime)$ is given by
\begin{eqnarray}
I(p^\prime,q,q^\prime)=\frac{1}{4(2\pi)^2}
\int_0^\infty \frac{{\bf |k|}}{\omega} d{\bf |k|}\int_0^{2\pi} 
d\phi\,\,
F_{\gamma}(\omega)\,\,F_e(E_p)\,\frac{1}{{\bf |P|}} 
\,\,\sum_{s,\epsilon} |{\mathcal{M}}|^2\,.
\label{Int}
\end{eqnarray}
The Legendre coefficients $\Phi^{\genfrac{}{}{0pt}{}{p}{a}}_l(E_q, E_{q^{\prime}})$ are determined from
\begin{eqnarray}\label{Legendre2}
\Phi^{\genfrac{}{}{0pt}{}{p}{a}}_l(E_q, E_{q^\prime}) 
&=&\int_{-1}^1 d(\cos\theta_{qq\prime}) P_l(\cos\theta_{qq\prime}) 
R^{\genfrac{}{}{0pt}{}{p}{a}}(E_q, E_{q^\prime}, \cos\theta_{qq\prime})\,.
\end{eqnarray}
Using Eq. (\ref{kerfinal}) in Eq. (\ref{Legendre2}), the  
Legendre coefficient for the production process can be expressed as

\begin{eqnarray}
\Phi^p_l(E_q, E_{q^\prime}) &=&
\frac{\pi^3}{(2\pi)^6 E_q E_{q^\prime}}
\int_{-1}^1 d(\cos\theta_{qq\prime}) P_l(\cos\theta_{qq\prime}) \int_0^\infty \frac{{\bf |p^\prime|}^{2}}
{E_{p^\prime}}d{\bf |p|}^\prime \int_{-1}^{1}
 d\cos\,\theta_{e}\int_0^{2\pi} 
d\phi_{e}[1-F_e(E_{p^\prime})]I(p^\prime,q, q^\prime)
\label{Legfinal} \,,\nonumber \\
\end{eqnarray}
where $I(p^\prime,q,q^\prime)$ is given by Eq. (\ref{Int}). 

\subsection*{Numerical results}
The structure of the squared matrix element allows us to decompose the
Legendre coefficients into parts that are symmetric and anti-symmetric
in the energies $E_q$ and $E_q^\prime$ of the two outgoing
neutrinos. For example, 
\begin{eqnarray}
	\Phi_l^{p}(E_q, E_{q^\prime})&=& 
	(C_V^2+C_A^2) \times 
		\{ \textrm{part symmetric in }(E_q,\, E_{q^\prime})\}
		\nonumber \\
	&&+(C_V^2-C_A^2)\times 
		\{ \textrm{part symmetric in }(E_q,\, E_{q^\prime})\}
		\nonumber \\
	&&+C_V C_A\times 
		\{ \textrm{part antisymmetric in }(E_q,\, E_{q^\prime})\}
	\, ,
\end{eqnarray}
where only the parts symmetric in $E_q$ and $E_q^\prime$ contribute to
the total emissivity $Q$ and rate $\Gamma$, since the anti-symmetric
part can be eliminated by the use of Lenard's identity
\cite{BEAUDET1, DICUS1, SCHINDER1}.  Furthermore, it is clear from
Eq. (\ref{RLegendre}) that only $\Phi^p_0(E_q, E_{q^\prime})$
contributes to the total emissivity $Q$ and rate $\Gamma$; terms with
$l \geq 1$ contribute only to the angular distribution.

\begin{figure}[ht!]
\begin{picture}(480, 250)(0, 0)
\put(050, 020){\includegraphics[height=200pt]{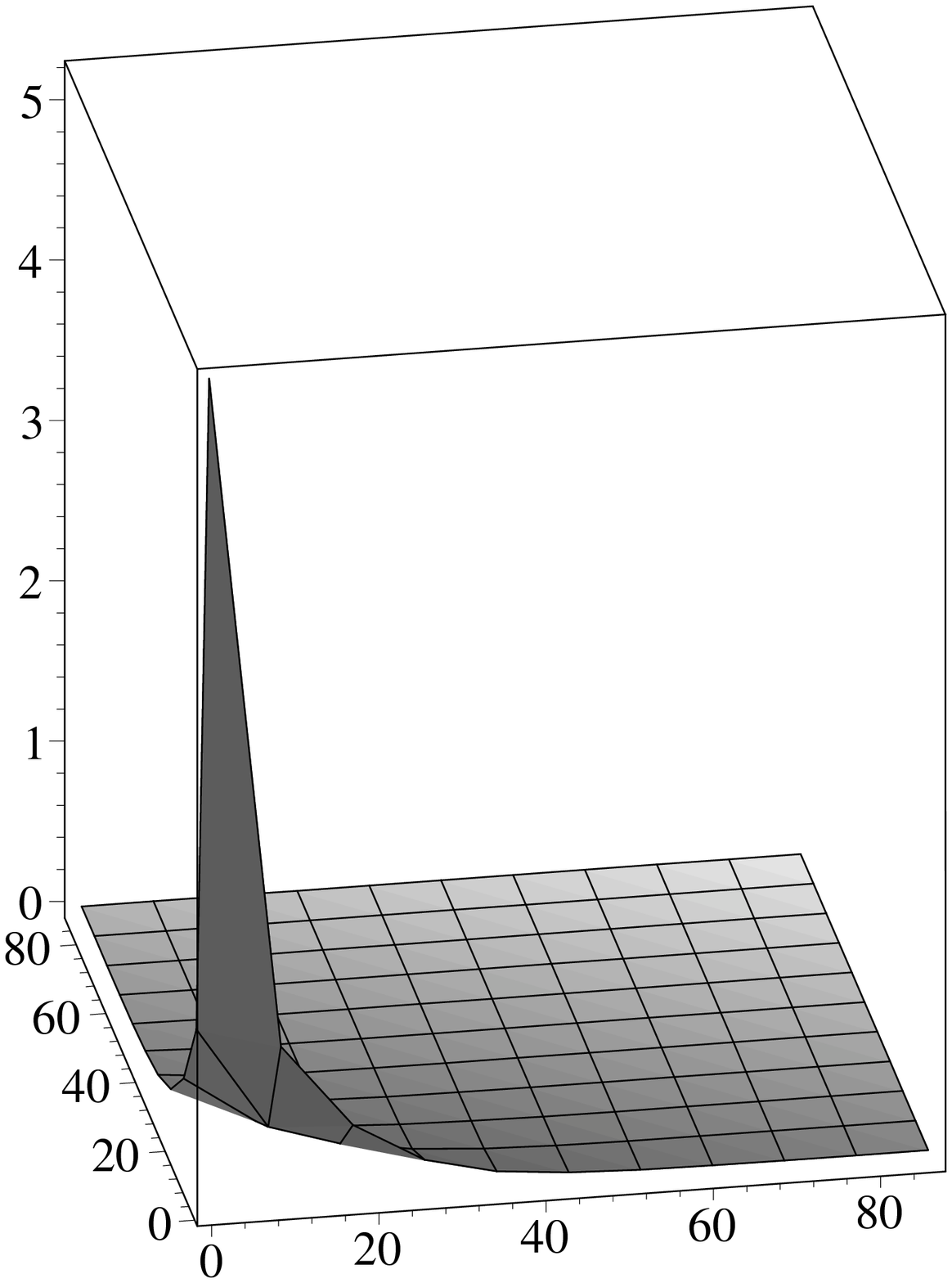}} \put(040,
040){\large $E_q$} \put(190, 010){\large $E_{q^\prime}$} \put(030,
220){\large $\Phi_0^{p\,({\rm sym})}$} \put(290,
020){\includegraphics[height=200pt]{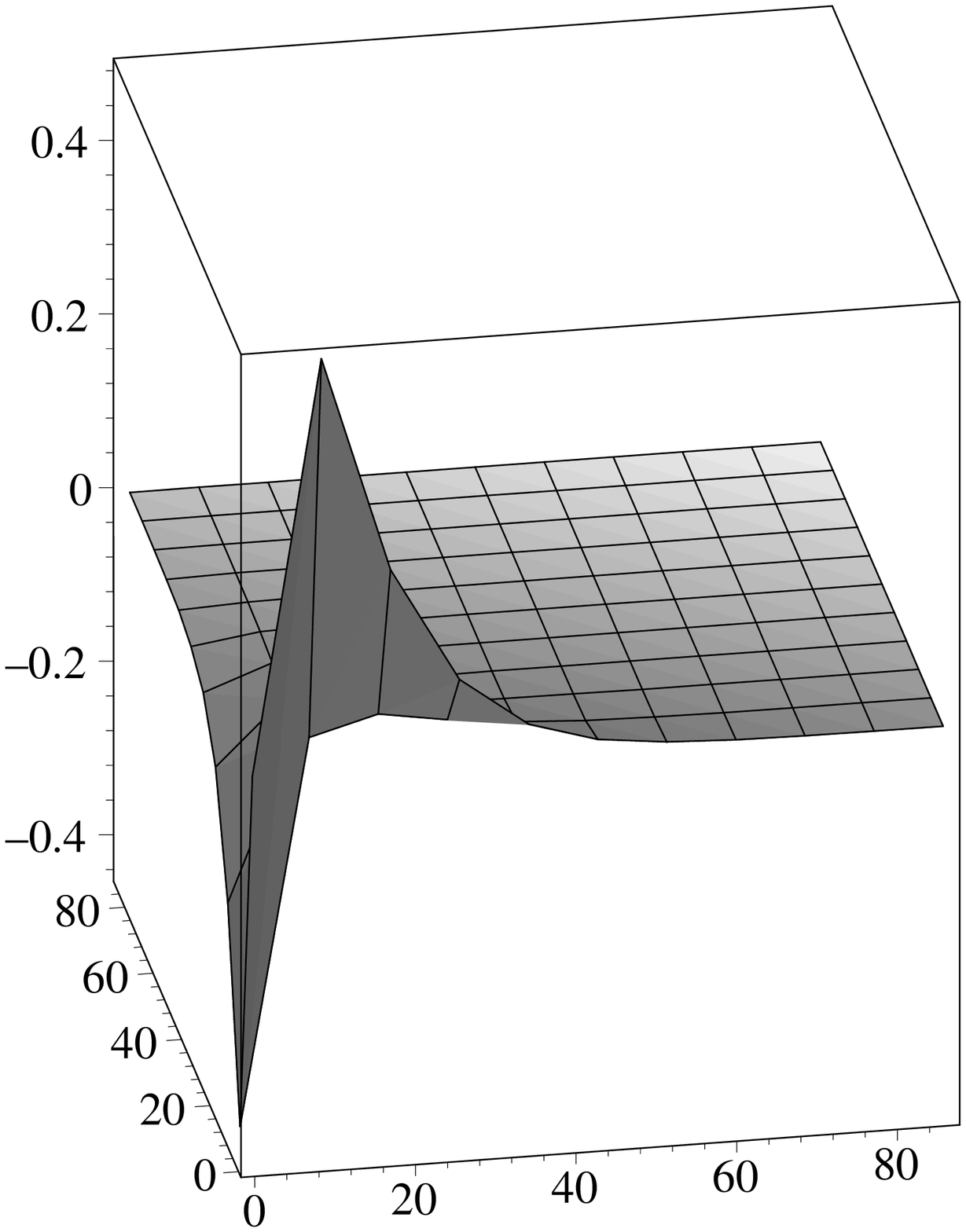}} \put(280, 040){\large
$E_q$} \put(430, 010){\large $E_{q^\prime}$} \put(270, 220){\large
$\Phi_0^{p\,({\rm asym})}$}
\end{picture}
\begin{picture}(480, 250)(0, 0)
\put(050, 020){\includegraphics[height=200pt]{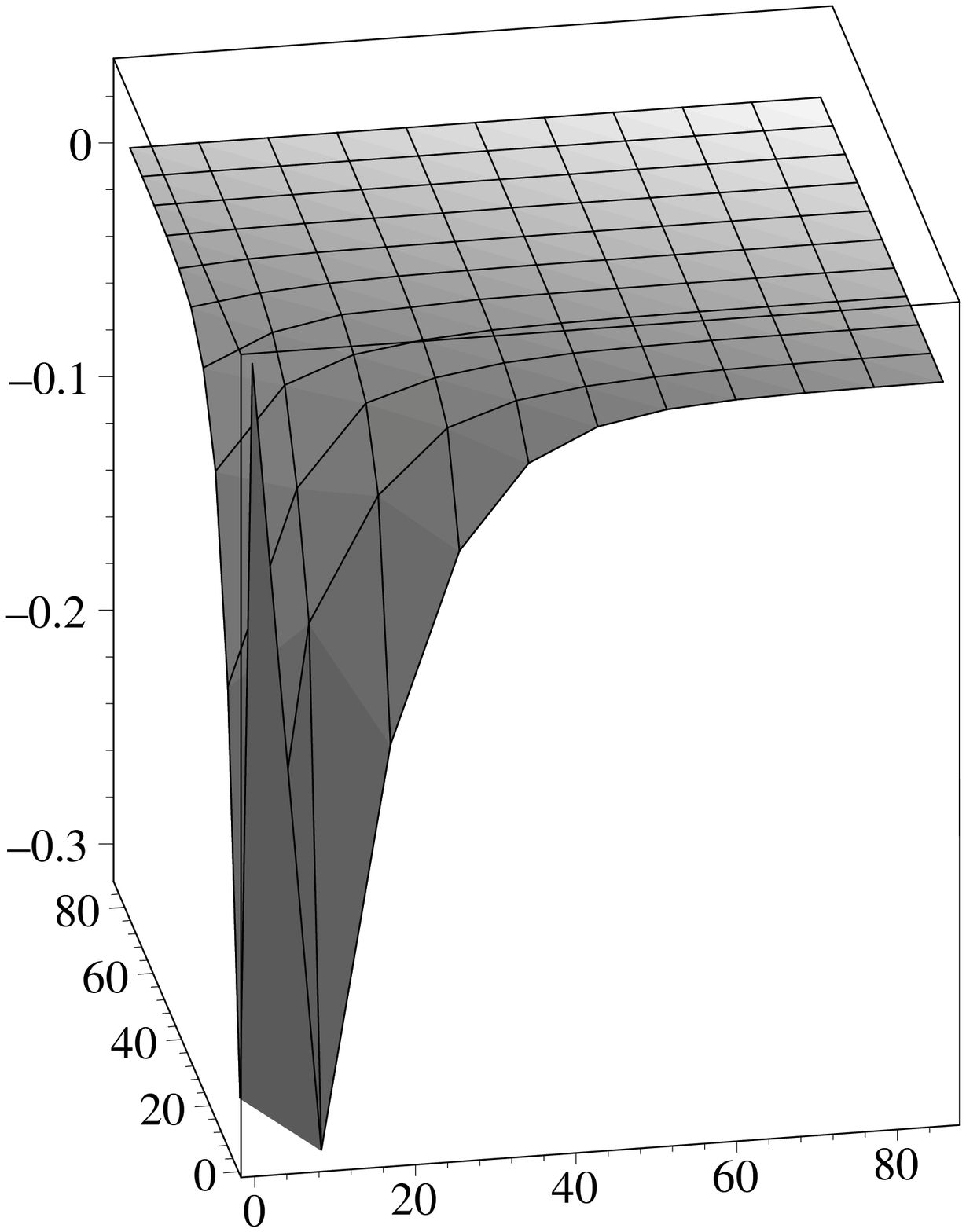}}
\put(040, 040){\large $E_q$}
\put(190, 010){\large $E_{q^\prime}$}
\put(030, 220){\large $\Phi_1^{p\,({\rm sym})}$}
\put(290, 020){\includegraphics[height=200pt]{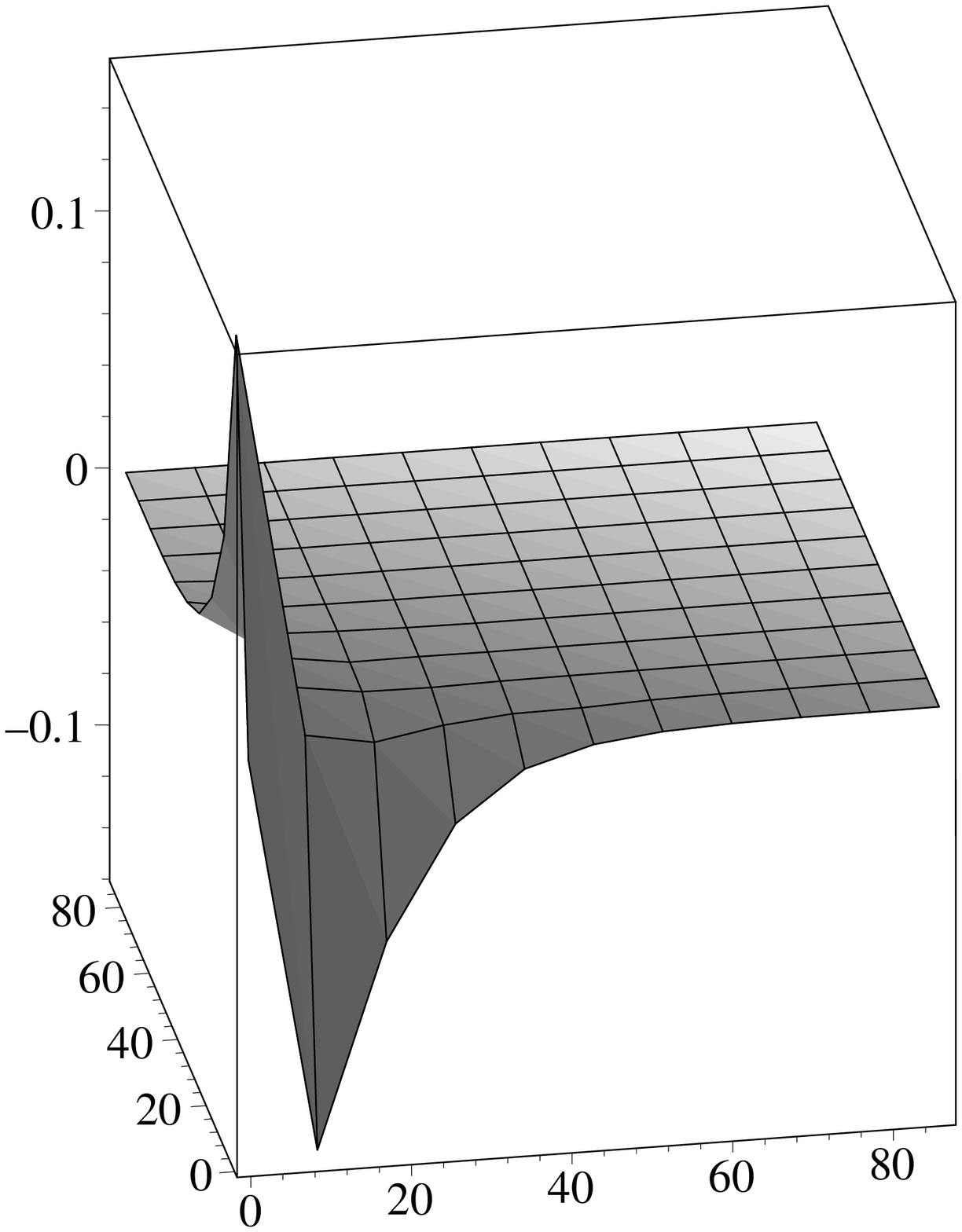}}
\put(280, 040){\large $E_q$}
\put(430, 010){\large $E_{q^\prime}$}
\put(270, 220){\large $\Phi_1^{p\,({\rm asym})}$}
\end{picture}
\caption{Symmetric and anti-symmetric parts of the Legendre
coefficients $\Phi_l^{p\,({\rm sym})}$ and $\Phi_l^{p\,({\rm asym})}$
in the production kernels for the
transverse case.  The neutrino energies $E_q$ and
$E_{q^\prime}$ are in MeV and the Legendre coefficients 
are in units of $10^{57}$ $\hbar^6$ erg$^{-6}$ cm$^3$ s$^{-7}$.
\label{fig:phikernels}
} 
\end{figure} 

In order to ascertain the importance of the $l \ge 1$ coefficients
relative to the $l = 0$ coefficient, numerical integrations of the
integrals in Eq. (\ref{Legfinal}) were performed for various values of
$l$ for the transverse case. Convergent numerical results were
obtained by using three different methods: (1) a Monte Carlo procedure
with uniform sampling using $10^5-10^6$ points, (2) the VEGAS Monte
Carlo \cite{NR} procedure that employs stratified importance sampling,
and (3) Gauss-Legendre quadrature with 8 and 16 points in each
dimension.  In regions of density and temperature where the resonant
character of the squared matrix element does not become prominent (see
Appendix B), the standard Monte Carlo and VEGAS procedures yield
similar variances ($\leq 5$\%).  However, a blind use of the VEGAS
routine from standard libraries fails to evaluate the integrals
properly in the high $T$ and $\rho_B$ regions in which the resonant
character of the squared matrix element strikes.  As discussed in
Appendix B, special care has to be exercised in treating this
situation.

Numerical results for the symmetric and anti-symmetric components of
$\Phi_0^p(E_q,E_{q^\prime})$ and $\Phi_1^p(E_q,E_{q^\prime})$ are
shown in Fig. \ref{fig:phikernels} for the case $T=10^{11}$ K$=8.62$
MeV and $\rho_BY_e=1$ g cm$^{-3}$.  The results explicitly show the
expected symmetry properties in $E_q$ and $E_{q^\prime}$.  A
comparison of the relative magnitudes in these four cases shows that
$\Phi_0^{p\, ({\rm sym})}(E_q,E_{q^\prime})$ is the dominant term.
The magnitude of $\Phi_0^{p\, ({\rm asym})}$ amounts to only 10\% of
the leading $\Phi_0^{p\, ({\rm sym})}$ contribution. The contributions
of $\Phi_1^{p\, ({\rm sym})}$ and $\Phi_1^{p\, ({\rm asym})}$ are 6\%
and 3\%, respectively.  In physical terms, this means that
neutrino-pair emission from the photo-neutrino process is dominantly
isotropic.  Therefore, depending on the required accuracy,
$\Phi_0^{p\, ({\rm sym})}$ might be adequate in practical
applications.  Note also that that the production kernels are
negligible for energies $E_q$ and $E_{q^\prime}$ $\gtrsim 10T$.

\section{Summary and Relation to Previous Works }
\label{sec:summary}
In summary, we have calculated the differential rates and emissivities
of neutrino pairs from the photo-neutrino process $e^\pm + \gamma
\rightarrow e^\pm + \nu + \bar\nu$ in an equilibrium plasma for widely
varying baryon densities and temperatures encountered in astrophysical
systems.  The new analytical expressions for the differential
emissivities yield total emissivities that are consistent with those
calculated in prior works
\cite{BEAUDET1,BEAUDET2,DICUS1,BOND1,SCHINDER1,ITOH}.  In order to
obtain these results, hitherto unavailable squared matrix elements for
this process were calculated for the transverse, longitudinal, and
mixed components.

We have developed new analytical expressions for the total
emissivities in various limiting situations.  These results help us to
better understand qualitatively, and in many cases quantitatively, the
scaling of the results with physical quantities such as the chemical
potential and plasma frequency at each temperature and density.

Using our results for the differential rates and emissivities, we have
calculated the production and absorption kernels in the source term of
the Boltzmann equation employed in exact, albeit numerical, treatments
of multienergy neutrino transport.  We have also provided the
appropriate Legendre coefficients of these kernels in forms suitable
for multigroup flux-limited diffusion schemes.

Beginning with the seminal work in Ref. \cite{BEAUDET1}, all prior
works \cite{BEAUDET2,DICUS1,BOND1,SCHINDER1,ITOH} have concentrated on
providing improved total rates and emissivities, but only for the
transverse component.  Our results for the total emissivity show that
the peak values of the longitudinal component are nearly equal to those
of the transverse component at all temperatures. (The mixed component
vanishes exactly for the total emissivity, but not for the
differential emissivity.)  Through the years, prior works have
persistently reported numerical difficulties in obtaining accurate
results at high temperatures and high density. In this work, the cause
of these difficulties has been traced to resonant factors that
misbehave when Monte Carlo techniques are employed to perform
multidimensional integrals. We have overcome these difficulties by
utilizing a principal value prescription.  
The prinicpal new elements of this work are
calculations of the full matrix elements, the neutrino differential rates and emissivities,  the
analytical analysis of the total emissivities in limiting situations, the
development of production and absorption kernels in the source term of
the Boltzmann equation for neutrino transport, and calculations of the
appropriate Legendre coefficients of these kernels in forms suitable
for multigroup flux-limited diffusion schemes.

\begin{acknowledgments}
We thank Doug Swesty, Jim Lattimer, and Eric Myra who alerted us to
the need for differential rates and emissivities in simulations of
neutrino transport in the supernova environment.  The work of of
S.I.D. was supported by the National Science Foundation Grant
No. 0070998 and by the Department of Energy grant DE-FG02-87ER-40317,
and that of S.R. and M.P. was supported by the US-DOE grant
DE-FG02-88ER40388.  Travel support for all three authors under the
cooperative agreement DE-FC02-01ER41185 for the SciDaC project
``Shedding New Light on Exploding Stars: Terascale Simulations of
Neutrino-Driven Supernovae and Their Nucleosynthesis'' is gratefully
acknowledged.

\end{acknowledgments}

\begin{appendix}
\section{Squared Matrix Elements}
\label{sec:smes}
The squared matrix element in Eq. (\ref{ME2}) contains the quantities 
$\mathcal{M_-},\,\mathcal{M_+}$ and $\mathcal{M_\times}$ which depend
on the scalar products of the various four momenta and the polarization
of the photon. The fruits of our toil are summarized by the following 
explicit expressions for these quantities.  
\begin{eqnarray}
\mathcal{M_-}&=&
-q\cdot q^{\prime}\biggl[k^2\epsilon\cdot\epsilon\,\,\mathcal{B_-}^2
+4\biggl(\frac{\epsilon\cdot p^2}{\beta_1^2}\,\,+\,\,
\frac{\epsilon\cdot p{^{\prime2}}}{\beta_2^2}+
\frac{2\epsilon \cdot  p\epsilon\cdot p^{\prime}}
{\beta_1\beta_2}\biggr)\biggr]
-4\biggl[\frac{k\cdot qk\cdot q^{\prime}\epsilon\cdot \epsilon + 
k^2\epsilon\cdot q\epsilon\cdot q^{\prime}}{\beta_1\beta_2}\biggr]
\end{eqnarray}
\begin{eqnarray}
\nonumber
\mathcal{M_+}&=&S(p,p^{\prime},q,q^{\prime})
\biggl[k^2\epsilon\cdot \epsilon\,\,\mathcal{B_+}^2
+4\biggl(\frac{\epsilon\cdot p^2}{\beta_1^2}+
\frac{\epsilon\cdot p{^{\prime 2}}}{\beta_2^2}+
\frac{2\epsilon\cdot p\epsilon\cdot p^{\prime}}{\beta_1\beta_2}\biggr)
\biggl]\\\nonumber&&
+2S(k,p^{\prime},q,q^{\prime})\biggl[\frac{2\epsilon\cdot p}{\beta_1}
\biggl(\frac{\epsilon\cdot p}{\beta_1}+
\frac{\epsilon\cdot p^{\prime}}{\beta_2}\biggl)-
\frac{k\cdot p\epsilon\cdot \epsilon}{\beta_1}\,\,\mathcal{B_+}
\biggr]
-2S(k,p,q,q^{\prime})
\biggl[\frac{2\epsilon\cdot p^{\prime}}{\beta_2}
\biggl(\frac{\epsilon\cdot p}{\beta_1} + 
\frac{\epsilon\cdot p^{\prime}}{\beta_2}\biggl)+
\frac{k\cdot p^{\prime}\epsilon\cdot \epsilon}{\beta_2}\,\,\mathcal{B_+}
\biggr]
\\\nonumber&&
+\frac{4}{\beta_1\beta_2}
\biggl[\epsilon\cdot p^{\prime}\epsilon\cdot q^{\prime}\,\,W(p,q,P_t)+
\epsilon\cdot p^{\prime}\epsilon\cdot q\,\,W(p,q^{\prime},P_t)
+\epsilon\cdot p\epsilon\cdot q^{\prime}\,\,W(p^{\prime},q,-P_t)
+\epsilon\cdot p\epsilon\cdot q\,\,W(p^{\prime},q^{\prime},-P_t)
 \\\nonumber&&
-\epsilon\cdot p^{\prime}\epsilon\cdot p(2k\cdot qk\cdot
q^{\prime}-k^2q\cdot q^{\prime})
+\epsilon\cdot q^{\prime}\epsilon\cdot q(k^2p\cdot p^{\prime}-2p\cdot
kp^{\prime} \cdot k) 
-\epsilon\cdot \epsilon\biggl(q\cdot q^{\prime}(k^2p\cdot p^{\prime}-p\cdot 
kp^{\prime}\cdot k)-p\cdot p^{\prime} k\cdot qk\cdot q^{\prime}\biggr)\biggr]\\
\end{eqnarray}
\begin{eqnarray}
\nonumber
{\mathcal{M_\times}}&=&
\frac{2}{\beta_1^2}
\biggl[(k^2 \epsilon\cdot\epsilon +4\epsilon\cdot p^2)\,R(p^{\prime},p,q,q^{\prime})
 + 2\epsilon \cdot p\,\beta_1\,(p^{\prime} \cdot q^{\prime}\epsilon \cdot q
-p^{\prime}  \cdot q\epsilon  
\cdot q^{\prime})
 +(-2p \cdot k\epsilon\cdot\epsilon+4\epsilon \cdot p^2)R(p^{\prime},k,q,q^{\prime})
\biggr]\\\nonumber&&
+\frac{2}{\beta_2^2}
\biggl[(k^2\epsilon\cdot\epsilon+4\epsilon \cdot p^{\prime 2})R(p^{\prime},p,q,q^{\prime})
 - 2\epsilon \cdot p^{\prime}\,\beta_2\,(p \cdot q^{\prime}\epsilon \cdot q
-p \cdot q\epsilon \cdot q^{\prime})
-(-2p^{\prime} \cdot k\epsilon\cdot\epsilon-4\epsilon \cdot p^{\prime 2})R(p,k,q,q^{\prime}) 
\biggr]\\\nonumber&&
+\frac{4}{\beta_1\beta_2}
Tr \biggl[2\epsilon \cdot p^{\prime}\epsilon \cdot p\biggl(R(p^{\prime},k,q,q^{\prime})+
2R(p^{\prime},p,q,q^{\prime})+
R(p,k,q,q^{\prime})\biggr)\\\nonumber&&
-\epsilon\cdot\epsilon
\biggl(k \cdot p^{\prime} R(p,k,q,q^{\prime})+
k^2R(p^{\prime},p,q,q^{\prime})-p \cdot kR(p^{\prime},k,q,q^{\prime})
\biggr)\\\nonumber&&
+\epsilon \cdot p\epsilon \cdot q \biggl
(2p^{\prime} \cdot kk \cdot q^{\prime}+2k \cdot p^{\prime} p \cdot q^{\prime}
-p^{\prime} \cdot q^{\prime} k^2\biggr)
-\epsilon \cdot p\epsilon \cdot q^{\prime}\biggl(2p^{\prime} \cdot k
k \cdot q+2k \cdot p^{\prime} p \cdot q
-p^{\prime} \cdot qk^2\biggr)\\&&
+\epsilon \cdot p^{\prime}\epsilon \cdot q^{\prime}
\biggl(2p \cdot k \, k \cdot q -2p^{\prime} \cdot qp \cdot k-p \cdot qk^2\biggr)
-\epsilon \cdot p^{\prime}\epsilon \cdot q
\biggl(2p \cdot k  \,k \cdot q^{\prime}-2p \cdot kp^{\prime} \cdot q^{\prime}
-p \cdot q^{\prime} k^2
\biggr)\biggr]
\end{eqnarray}
In writing the above expressions, we have employed 
the following  notations: 
\begin{eqnarray}
P_t = p+p^{\prime}\,, \qquad
{\mathcal{B_-}} &=& \frac{1}{\beta_1}-\frac{1}{\beta_2}\,, \qquad
{\mathcal{B_+}} = \frac{1}{\beta_1}+\frac{1}{\beta_2}\,, \nonumber \\\nonumber\\
S(x,y,w,z)&=&(x\cdot w)\,\, (y\cdot z)+(x\cdot z) \,\,(y\cdot w)\,, \nonumber \\
R(x,y,w,z)&=&(x\cdot w)\,\, (y\cdot z)-(x\cdot z) \,\,(y\cdot w)\,, 
\quad {\rm and} \nonumber \\
W(x,y,z)&=&(x\cdot k)\,\,( y\cdot k )-(x\cdot y)\,\,
k^2-(x\cdot k)\,\,(z\cdot y) \, .
\end{eqnarray}

After summing over the polarization of the photon, the transverse and
longitudinal components of the quantities
${\mathcal{M_-}}^{T(L)},\,{\mathcal{M_+}}^{T(L)}$ and $
{\mathcal{M_\times}}^{T(L)}$ required in the calculation of the
squared matrix elements in Eq. (\ref{ME2TL}) are
\begin{eqnarray}
{\mathcal{M_-}}^T&=&
2q\cdot q^{\prime}\biggl[k^2\,\,\mathcal{B_-}^2
-2\biggl(\frac{P^T(p,p)}{\beta_1^2}\,\,+\,\,
\frac{P^T(p^{\prime},p^{\prime})}{\beta_2^2}+
\frac{2P^T(p,p^{\prime})}
{\beta_1\beta_2}\biggr)\biggr]
+4\biggl[\frac{2k\cdot qk\cdot q^{\prime}-k^2
P^T(q,q^{\prime})}{\beta_1\beta_2}\biggr]
\end{eqnarray}
\begin{eqnarray}
\nonumber
{\mathcal{M_+}}^T&=&2S(p,p^{\prime},q,q^{\prime})
\biggl[-k^2\,\,\mathcal{B_+}^2
+2\biggl(\frac{P^T(p,p)}{\beta_1^2}\,\,+\,\,
\frac{P^T(p^{\prime},p^{\prime})}{\beta_2^2}+
\frac{2P^T(p,p^{\prime})}
{\beta_1\beta_2}\biggr)
\biggl]\\\nonumber&&
+4S(k,p^{\prime},q,q^{\prime})\biggl[\frac{k\cdot p}{\beta_1}\,\,\mathcal{B_+}+
\frac{1}{\beta_1}
\biggl(\frac{P^T(p,p)}{\beta_1}+\frac{P^T(p,p^{\prime})}{\beta_2}\biggl)
\biggr]\\\nonumber&&
+4S(k,p,q,q^{\prime})
\biggl[\frac{k\cdot p^{\prime}}{\beta_2}\,\,\mathcal{B_+}
-\frac{1}{\beta_2}
\biggl(\frac{P^T(p,p^{\prime})}{\beta_1}+\frac{P^T(p^{\prime},p^{\prime})}{\beta_2}\biggl)
\biggr]
\\\nonumber&&
+\frac{4}{\beta_1\beta_2}
\biggl[P^T(p^{\prime},q^{\prime})\,W(p,q,P_t)+
P^T(p^{\prime},q)\,W(p,q^{\prime},P_t)
+P^T(p,q^{\prime})\,W(p^{\prime},q,-P_t)
+P^T(p,q)\,W(p^{\prime},q^{\prime},-P_t)
\\\nonumber&&
-P^T(p,p^{\prime})(2k\cdot qk\cdot q^{\prime}-k^2q\cdot q^{\prime})
+P^T(q,q^{\prime})(k^2p\cdot p^{\prime}-2p\cdot kp^{\prime}\cdot k)
+2q\cdot q^{\prime}(k^2p\cdot p^{\prime}-p\cdot kp^{\prime}\cdot
k)-2p\cdot p^{\prime} k\cdot qk\cdot q^{\prime}\biggr]
\,.\\\\\nonumber
\end{eqnarray}
\begin{eqnarray}
\nonumber
{\mathcal{M_\times}}^T&=& \frac{2}{\beta_1^2}\biggl\{
\biggl(-2k^2+4P_T(p,p)\biggr)\,R(p^{\prime},p,q,q^{\prime}) + 2\beta_1\biggl(
p^{\prime}\cdot q^{\prime} P_T(p,q)-p^{\prime}\cdot q P_T(p,q^{\prime})\biggr)\\\nonumber&&
+\biggl(4p\cdot k+4P_T(p,p)\biggr)R(p^{\prime},k,q,q^{\prime})
\biggr\}\\\nonumber&& +\frac{2}{\beta_2^2}\biggl\{
\biggl(-2k^2+4P_T(p^{\prime},p^{\prime})\biggr)R(p^{\prime},p,q,q^{\prime}) - 2\beta_2
\biggl(p\cdot q^{\prime} P_T(p^{\prime},q)- p\cdot q
P_T(p^{\prime},q^{\prime})\biggr)\\\nonumber&&
-\biggl(4p^{\prime}\cdot k-4P_T(p^{\prime},p^{\prime})\biggr)R(p,k,q,q^{\prime})
\biggr\}\\\nonumber&& +\frac{4}{\beta_1\beta_2}\biggl\{
2P_T(p,p^{\prime})\biggl(R(p^{\prime},k,q,q^{\prime})+2R(p^{\prime},p,q,q^{\prime})+
R(p,k,q,q^{\prime})\biggr)\\\nonumber&&
+2\biggl(k\cdot p^{\prime} R(p,k,q,q^{\prime})+k^2R(p^{\prime},p,q,q^{\prime})
-p\cdot kR(p^{\prime},k,q,q^{\prime})\biggr)\\\nonumber&&
+P_T(p,q)\biggl(2p^{\prime}\cdot kk\cdot q^{\prime}+2k\cdot p^{\prime} p\cdot q^{\prime}-p^{\prime}\cdot q^{\prime}
k^2\biggr)
-P_T(p,q^{\prime})\biggl(2p^{\prime}\cdot kk\cdot q+2k\cdot p^{\prime} p\cdot q
-p^{\prime}\cdot qk^2\biggr)\\&& +P_T(p^{\prime},q^{\prime})\biggl(2p\cdot k \, k\cdot q
-2p^{\prime}\cdot qp\cdot k-p\cdot qk^2\biggr) -P_T(p^{\prime},q)\biggl(2p\cdot k
\,k\cdot q^{\prime}-2p\cdot kp^{\prime}\cdot q^{\prime}-p\cdot q^{\prime} k^2 \biggr)\biggr\}
\end{eqnarray}
The corresponding expressions for the longitudinal component are
\begin{eqnarray}
{\mathcal{M_-}}^L&=& q\cdot q^{\prime}\biggl[k^2\,\,\mathcal{B_-}^2
-4\biggl(\frac{P^L(p,p)}{\beta_1^2}\,\,+\,\,
\frac{P^L(p^{\prime},p^{\prime})}{\beta_2^2}+ \frac{2P^L(p,p^{\prime})}
{\beta_1\beta_2}\biggr)\biggr] +4\biggl[\frac{k\cdot qk\cdot
q^{\prime}-k^2
P^L(q,q^{\prime})}{\beta_1\beta_2}\biggr]
\end{eqnarray}
\begin{eqnarray}
\nonumber
{\mathcal{M_+}}^L&=&S(p,p^{\prime},q,q^{\prime})
\biggl[-k^2\,\,\mathcal{B_+}^2
+4\biggl(\frac{P^L(p,p)}{\beta_1^2}\,\,+\,\,
\frac{P^L(p^{\prime},p^{\prime})}{\beta_2^2}+
\frac{2P^L(p,p^{\prime})}
{\beta_1\beta_2}\biggr)
\biggl]\\\nonumber&&
+2S(k,p^{\prime},q,q^{\prime})\biggl[\frac{k\cdot p}{\beta_1}\,\,\mathcal{B_+}+
\frac{2}{\beta_1}
\biggl(\frac{P^L(p,p)}{\beta_1}+\frac{P^L(p,p^{\prime})}{\beta_2}\biggl)
\biggr]\\\nonumber&&
+2S(k,p,q,q^{\prime})
\biggl[\frac{k\cdot p^{\prime}}{\beta_2}\,\,\mathcal{B_+}
-\frac{2}{\beta_2}
\biggl(\frac{P^L(p,p^{\prime})}{\beta_1}+\frac{P^L(p^{\prime},p^{\prime})}{\beta_2}\biggl)
\biggr]
\\\nonumber&&
+\frac{4}{\beta_1\beta_2}
\biggl[P^L(p^{\prime},q^{\prime})\,W(p,q,P_t)+
P^L(p^{\prime},q)\,W(p,q^{\prime},P_t)
+P^L(p,q^{\prime})\,W(p^{\prime},q,-P_t)
+P^L(p,q)\,W(p^{\prime},q^{\prime},-P_t)
\\\nonumber&&
-P^L(p,p^{\prime})(2k\cdot qk\cdot q^{\prime}-k^2q\cdot q^{\prime})
+P^L(q,q^{\prime})(k^2p\cdot p^{\prime}-2p\cdot kp^{\prime}\cdot k)
\\&&
+q\cdot q^{\prime}(k^2p\cdot p^{\prime}-p\cdot kp^{\prime}\cdot
k)-p\cdot p^{\prime} k\cdot qk\cdot q^{\prime}\biggr].
\end{eqnarray}
\begin{eqnarray}
\nonumber {\mathcal{M_\times}}^L &=& \frac{2}{\beta_1^2}\biggl\{
\biggl(-k^2+4P_L(p,p)\biggr)\,R(p^{\prime},p,q,q^{\prime}) + 2\beta_1\biggl(
p^{\prime}\cdot q^{\prime} P_L(p,q)-p^{\prime}\cdot q
P_L(p,q^{\prime})\biggr)
\\\nonumber&&
+\biggl(2p\cdot k+4P_L(p,p)\biggr)R(p^{\prime},k,q,q^{\prime})
\biggr\}\\\nonumber&& +\frac{2}{\beta_2^2}\biggl\{
\biggl(-k^2+4P_L(p^{\prime},p^{\prime})\biggr)R(p^{\prime},p,q,q^{\prime}) 
- 2\beta_2
\biggl(p\cdot q^{\prime} P_L(p^{\prime},q)- 
p\cdot q P_L(p^{\prime},q^{\prime})\biggr)
\\\nonumber&&
- \biggl(2p^{\prime}\cdot k-4P_L(p^{\prime},p^{\prime})\biggr)
R(p,k,q,q^{\prime})
\biggr\}\\\nonumber&& +\frac{4}{\beta_1\beta_2}\biggl\{
2P_L(p,p^{\prime})\biggl(R(p^{\prime},k,q,q^{\prime})+2R(p^{\prime},p,q,q^{\prime})+
R(p,k,q,q^{\prime})\biggr)\\\nonumber&&
+\biggl(k\cdot p^{\prime} R(p,k,q,q^{\prime})+
k^2R(p^{\prime},p,q,q^{\prime})-p\cdot kR(p^{\prime},k,q,q^{\prime})\biggr)\\\nonumber&&
+P_L(p,q)\biggl(2p^{\prime}\cdot kk\cdot q^{\prime}+2k\cdot p^{\prime} p\cdot q^{\prime}-p^{\prime}\cdot q^{\prime} k^2\biggr)
-P_L(p,q^{\prime})\biggl(2p^{\prime}\cdot kk\cdot q+2k\cdot p^{\prime} p\cdot q
-p^{\prime}\cdot qk^2\biggr)\\&& +P_L(p^{\prime},q^{\prime})\biggl(2p\cdot k \, k\cdot q
-2p^{\prime}\cdot qp\cdot k-p\cdot qk^2\biggr) -P_L(p^{\prime},q)\biggl(2p\cdot k
\,k\cdot q^{\prime}-2p\cdot kp^{\prime}\cdot q^{\prime}-p\cdot q^{\prime} k^2 \biggr)\biggr\}
\end{eqnarray}
While summing over the photon polarizations, we have utilized the 
relations 
\begin{eqnarray}
P^T(x,y) &=& {\bf x}\cdot{\bf y} - \frac{({\bf x}\cdot{\bf k})\,\, ({\bf y}
\cdot {\bf k})}{{\bf k}^2} \,, \nonumber \\
P^L(x,y) &=& - x\cdot y+\frac{(x\cdot k)\,\, (y \cdot k)}{k^2}
-P^T(x,y) \,.
\end{eqnarray}

For completeness, the trace of six $\gamma$--matrices employed in the
calculations above is given below:
\begin{eqnarray}
Tr[\gamma_\mu \gamma_\nu \gamma_\sigma \gamma_\tau \gamma_\alpha
  \gamma_\beta]
  &=&
	\, 4 g_{\alpha\beta} \big( g_{\mu\nu}g_{\sigma\tau} - 
g_{\mu\sigma}g_{\nu\tau} + g_{\mu\tau}g_{\nu\sigma} \big) \nonumber\\
	&&-4 g_{\alpha\mu} \big( g_{\nu\sigma}g_{\tau\beta} - 
g_{\sigma\beta}g_{\nu\tau} + g_{\sigma\tau}g_{\nu\beta} \big) \nonumber\\
	&&+4 g_{\alpha\nu} \big( g_{\mu\sigma}g_{\tau\beta} - 
g_{\mu\tau}g_{\sigma\beta} + g_{\mu\beta}g_{\sigma\tau} \big) \nonumber\\
	&&-4 g_{\alpha\sigma} \big( g_{\mu\nu}g_{\tau\beta} - 
g_{\mu\tau}g_{\nu\beta} + g_{\mu\beta}g_{\nu\tau} \big) \nonumber\\
	&&+4 g_{\alpha\tau} \big( g_{\mu\nu}g_{\sigma\beta} - 
g_{\mu\sigma}g_{\nu\beta} + g_{\mu\beta}g_{\nu\sigma} \big) \, .
\end{eqnarray}
The trace with an additional $\gamma_5$ was computed by using
\begin{eqnarray}
Tr[\gamma_\mu \gamma_\nu \gamma_\sigma \gamma_\tau \gamma_\alpha 
\gamma_\beta \gamma_5]  &=&
	4 i \big(g_{\mu\nu}\epsilon_{\sigma\tau\alpha\beta} - 
g_{\mu\sigma}\epsilon_{\nu\tau\alpha\beta} + 
g_{\nu\sigma}\epsilon_{\mu\tau\alpha\beta} \nonumber\\
		&&+ g_{\alpha\tau}\epsilon_{\mu\nu\sigma\beta} - 
g_{\tau\beta}\epsilon_{\mu\nu\sigma\alpha} + 
g_{\alpha\beta}\epsilon_{\mu\nu\sigma\tau}\big) \, .
\end{eqnarray}

\section{Treatment of Resonant Factors}
\label{sec:singu}

In the high temperature regime, $T\gtrsim m_e$, and for densities in
the range $10^{10} \lsim \rho_B Y_e/{\rm g~ cm^{-3}} \lsim 10^{12}$, prior
works~ \cite{BEAUDET1,BEAUDET2,DICUS1,BOND1,SCHINDER1,ITOH} have
reported numerical problems in calculating the neutrino emissivities
through the use of Monte Carlo procedures.  Our purpose here is to
offer a solution to this longstanding problem.

We have traced the origin of this problem to the factor
\begin{equation}
  \frac{1}{\beta_2}=\frac{1}{k^2-2p{^\prime}\cdot k} = 
 \frac {1} {k^2 - 2 E_{p^{\prime}} \omega 
+ 2 {\bf|k ||p^{\prime}|}\cos [\sphericalangle({\bf k},{\bf
 p^{\prime}})]} 
\label{dbeta2} 
\end{equation} 
and its square in the squared matrix element. This factor gives rise to a
sharp peak in the integrand for certain physical conditions in the
plasma.  In performing Monte Carlo integrations of Eq. (\ref{Int8}),
this feature causes a spurious increase in the emissivity as well
as in its variance.

At high temperatures, this numerical problem begins to occur at low
densities, peaks at intermediate densities (in the case that
$\omega_p<T<\mu_e$), and vanishes at sufficiently high densities.  In
what follows, we analyze the transverse component for which we use the
lowest order dispersion relation.  Generalization to the longitudinal
component and/or the full dispersion relation is straightforward.  The
cases of $T<m_e$ and $T>m_e$ have distinctly different behaviors and
are therefore considered separately.

We begin by enquiring why this problem does not occur in the low
temperature and low density regime ($T\ll m_e$ and $\mu_e \ll T$).  In
this case, the energy of the outgoing electron is
$E_{p^{\prime}}\simeq m_e$ and its momentum ${\bf |p^{\prime}|}\simeq
0$. Furthermore, $\omega_p \ll m_e$ (see Fig. \ref{WPMUFIGURE}).
Therefore, we can drop the first and last terms in the denominator of 
Eq. (\ref{dbeta2}) so that
\begin{equation}
\frac {1}{\beta_2} \simeq \frac{1} {- 2E_p{^{\prime}} \omega}  
\end{equation} 
which is perfectly regular.

With increasing density, the plasma begins to enter into the
degenerate regime in which $\mu_e\gsim T $, whereas $\mu_e\lesssim
m_e$. In this case, the energy and momentum of the outgoing electron
can be approximated by $E_{p^{\prime}}\simeq{\bf
|p^{\prime}|}\simeq\mu_e$.  For $\omega_p \gg \mu_e$, significant
contributions arise from $\omega\simeq\omega_p$ and ${\bf| k |}\simeq
0$ only, since photons with high energies/momenta are exponentially
suppressed by the tail of the Bose-Einstein distribution
function.  In this regime, therefore, either ${\bf |p^{\prime}|}$ or
${\bf| k |}$ is small.  This allows us to drop the last term in the
denominator of Eq. (\ref{dbeta2}) so that
\begin{equation}
\frac {1}{\beta_2} \simeq \frac{1}{\omega_p^2 - 2 \mu_e \omega_p}  
\end{equation}
which is regular, since $\mu_e \gg \omega_p$. 
In physical terms, the fact that the energy of the outgoing
electron is much larger than that of the photon ensures that
Eq. (\ref{dbeta2}) remains regular.

However, when the temperature exceeds the electron mass ($T\ge m_e$),
Eq. (\ref{dbeta2}) can become extremely large. At low
densities, which corresponds to the nondegenerate case ($T\gg
\mu_e$), $\omega_p > m_e$ (see Fig.  \ref{WPMUFIGURE}).  In this regime, we
can ignore neither the momentum of the electron compared to its energy
nor the energy of the photon compared to that of the electron. The
physical conditions are thus ripe for Eq. (\ref{dbeta2}) to blow
up. The worst happens for intermediate densities for which $\omega_p
\ll T \ll \mu_e$.  In this case, we can ignore the plasma frequency, 
set $E_{p^{\prime}}\simeq {\bf |p^{\prime}|}\simeq \mu_e$ and
$\omega\simeq {\bf |k|}$ so that Eq. (\ref{dbeta2}) takes the form
\begin{equation}
\frac{1}{\beta_2} \simeq \frac{1}{-2\mu_e\omega(1+
	\cos [\sphericalangle({\bf k},{\bf p^{\prime}})])}
\label{prob}
\end{equation}
which becomes large when the angle between ${\bf k}$ and $\bf
{p}^{\prime}$ approaches $\pi$.

As the density increases further, $\omega_p \gg {\bf|k|}$ (see the all
important Fig. \ref{WPMUFIGURE} once more) and hence we can set
$\omega=\omega_p$ for all relevant regions of the phase space not
supressed by the Bose-Einstein distribution function. Since in this region, we
have $E_{p^{\prime}}\simeq {\bf |p^{\prime}|}\simeq \mu_e\gg
\omega_p$, Eq. (\ref{dbeta2}) turns regular again.

In order to avoid the numerical problem associated with
Eq. (\ref{prob}), it is necessary to tame the resonant character
within the context of a Monte Carlo integration. Since we are
evaluating a multidimensional integral, it is cumbersome to apply a
different choice of variable sampling because the sampling function itself
would depend on the remaining variables.  The solution that is
computationally simple is to
evaluate the Cauchy principal value of the integral (see, for example,
Ref. \cite{DavisRabinowitz}). This can be achieved simply by excluding
the points when the angle between ${\bf k}$ and ${\bf p}^{\prime}$ is
within ($\alpha -\epsilon,\, \alpha +\epsilon$), where $\alpha$ is
given by
\begin{equation}
	\cos \alpha = \frac{2E_{p^{\prime}}\omega-k^2}{
2{\bf|k ||p^{\prime}|}}\, .
\end{equation}
In our computations, $\epsilon\simeq 0.01$ proved to be effective in
reducing the variance
significantly. Note that this procedure can be adopted for any of the
integration variables in Eq. (\ref{dbeta2}), but at the expense of
locating the lurking problems therein. 

We note, however that at high temperatures and high density, the plasmon
and pair-neutrino neutrino processes generally dominate over the
photo-neutrino process.

\end{appendix}

\end{document}